\documentclass[a4paper,12pt]{article}

\usepackage[margin=2.5cm]{geometry}
\usepackage{graphicx}
\usepackage{amssymb}
\usepackage{multirow}
\usepackage{amsmath}
\usepackage{setspace}
\usepackage{caption}
\usepackage{subcaption}
\usepackage[colorlinks=true,bookmarks=true]{hyperref}
\usepackage{float}
\usepackage{mathrsfs}
\usepackage[nosort]{cite}

\hypersetup{citecolor=blue}
\newcommand{\dif}{\mathrm{d}}

\newcommand{\Eqref}[1]{(\ref{#1})}
\newcommand{\half}{\frac{1}{2}}

\newcommand{\brac}[1]{\left(#1 \right)}
\newcommand{\sbrac}[1]{\left[#1\right]}

\newcommand{\eq}{\,=\,}

\makeatletter
\renewcommand\section{\@startsection {section}{1}{\z@}%
                                   {-3.5ex \@plus -1ex \@minus -.2ex}%
                                   {2.3ex \@plus.2ex}%
                                   {\normalfont\bfseries}}
\renewcommand\subsection{\@startsection{subsection}{2}{\z@}%
                                     {-3.25ex\@plus -1ex \@minus -.2ex}%
                                     {1.5ex \@plus .2ex}%
                                     {\normalfont\it}}
\renewcommand\subsubsection{\@startsection{subsubsection}{3}{\z@}%
                                     {-3.25ex\@plus -1ex \@minus -.2ex}%
                                     {1.5ex \@plus .2ex}%
                                     {\normalfont}}
\makeatother

\begin{document}

\thispagestyle{empty}
\begin{flushright}

\end{flushright}
\vbox{}
\vspace{2cm}

\begin{center}
{\LARGE{Deformed hyperbolic black holes
        }}\\[16mm]
{{Yu Chen,~~Yen-Kheng Lim~~and~~Edward Teo}}
\\[6mm]
{\it Department of Physics,
National University of Singapore, 
Singapore 119260}\\[15mm]

\end{center}
\vspace{2cm}

\centerline{\bf Abstract}
\bigskip
\noindent
Black holes with planar or hyperbolic horizons are known to exist in AdS space, alongside the usual ones with spherical horizons. In this paper, we consider a one-parameter generalisation of these black holes that is contained in the AdS C-metric. In terms of the domain-structure analysis recently developed for such solutions, these black holes have a domain in the shape of a triangle. It is shown that the horizons of these black holes are deformed hyperbolic spaces, with the new parameter controlling the amount of deformation. The space-times are static and completely regular outside the horizons. We argue that these black holes are hyperbolic analogues of the ``slowly accelerating'' spherical black holes known to exist in AdS space.

\newpage

\section{Introduction}

Although our universe does not appear to have a negative cosmological constant, the study of anti-de Sitter (AdS) spaces has garnered much attention since the advent of the AdS/CFT correspondence (see, e.g., \cite{Hubeny:2014bla} for a recent review). This correspondence allows one to use an AdS space in general relativity to provide a holographic description of some strongly coupled quantum field theory. Moreover, if this field theory is to be at finite temperature, there is typically a black hole in the AdS space. This has led to AdS black holes being used to study problems in fields as diverse as nuclear physics and condensed matter physics (see, e.g., \cite{Janik:2010we,Musso:2014efa,Cai:2015cya}). Thus AdS black holes have been of great theoretical interest, and it is important to build up our repertoire and understanding of such solutions.

It is somewhat fortunate that there is no lack of known black-hole solutions in AdS space. Most of the Ricci-flat black holes, such as Schwarzschild and Kerr, have AdS analogues. Besides, there are known examples of AdS black holes with {\it no\/} Ricci-flat analogues. One such class are the so-called topological black holes \cite{Lemos:1994xp,Lemos:1994fn,Lemos:1995cm,Cai:1996eg,Mann:1996gj,Vanzo:1997gw,Brill:1997mf,Mann:1997iz,Birmingham:1998nr}. These are black holes whose horizons can either be spherical, planar or hyperbolic, depending on the value of a discrete parameter in the solution. In the spherical case, this is just the usual Schwarzschild-AdS black hole. In the planar and hyperbolic cases, the horizons are non-compact, but they can be made compact under appropriate identifications. Their topology will then become either that of a torus (for the planar case) or some Riemann surface with higher genus (for the hyperbolic case).\footnote{In this paper however, we shall not be performing these identifications, so the horizons will remain non-compact; nevertheless we shall continue to refer to them collectively as topological black holes.} Such black holes, with non-spherical horizon topology, do not exist when the cosmological constant is zero or positive. 

Since the Schwarzschild-AdS black hole can be generalised in several ways, it is natural to wonder if the above topological black holes can be similarly generalised. For example, one might ask if it is possible to include a rotation; indeed, a rotating hyperbolic black hole was found a number of years ago \cite{Klemm:1997ea}, by a suitable analytic continuation of the Kerr-AdS black hole. Another possible generalisation of the  Schwarzschild-AdS black hole is to include an acceleration. Such an accelerating black-hole solution is described by the well-known C-metric, in this case specifically the AdS C-metric. By appropriately adjusting some of the parameters of this solution, it is known to contain the topological black holes in the zero-acceleration limit \cite{Mann:1996gj}. 

Such accelerating topological black holes have been studied in the past, but mostly with a view for applications. For example, they have been used to calculate the pair production rate of topological black holes from the vacuum \cite{Mann:1996gj}, and to construct localised brane-world black holes \cite{Emparan:1999fd}. More recently, these solutions have been used to provide examples of so-called black funnels and black droplets \cite{Hubeny:2009ru,Hubeny:2009kz}, which describe strongly coupled field theories in black-hole backgrounds via the AdS/CFT correspondence. In particular, the structure of these solutions was analysed in some detail in \cite{Hubeny:2009kz}. However, it might still be worthwhile to study these solutions in their own right, to understand their geometries and physical interpretation, as well as how they are related to the usual accelerating spherical black holes.

In a previous paper \cite{Chen:2015vma}, we found a new form of the C-metric with cosmological constant. This form, which is given in (\ref{Cmetric_alphabeta}) below, is novel in the sense that its two (cubic) structure functions are each partially factorised in terms of two real roots. The solution is then parameterised in terms of the four roots. These roots are significant in that they define the boundaries of the range for two of the coordinates. This leads to a natural representation of the solution when the coordinate range is plotted out: it will fill out a rectangle or ``box'' in a two-dimensional plot. The locations of the edges of this box are given by the four roots.

In \cite{Chen:2015vma}, we referred to this box as the domain of the space-time. Furthermore, we showed how the shape of the domain can be used to deduce key physical information about the space-time. In this case, two opposite edges of the box actually correspond to axes in the space-time, while the other two opposite edges correspond to horizons in the space-time. From this, we can conclude that the space-time contains a spherical black hole undergoing an acceleration, with an acceleration horizon present. This is consistent with the usual physical interpretation of the C-metric.

In the case of negative cosmological constant, it turns out that the box is cut off at one corner by a diagonal line, which physically represents asymptotic infinity. While this does not change the above physical interpretation of the space-time, it does lead to the possibility of other types of domains in this case. For example, the corner that is cut off is a triangular domain, which corresponds to a space-time with only one axis and one horizon. Moreover, the axis and horizon both extend to infinity, and so are non-compact. It turns out that such domains describe accelerating generalisations of the planar and hyperbolic black holes of \cite{Lemos:1994xp,Lemos:1994fn,Lemos:1995cm,Cai:1996eg,Mann:1996gj,Vanzo:1997gw,Brill:1997mf,Mann:1997iz,Birmingham:1998nr}.

It is the aim of this paper to study the space-times described by these triangular domains in more detail. But before we do so, we need a slightly more general form of the AdS C-metric than that used in \cite{Chen:2015vma}. As mentioned above, the form used there assumes the existence of two real roots for each structure function. This was motivated by our primary aim in that paper to study the space-times described by the box-like domains. However, for triangular domains, it is only necessary to assume the existence of one real root for each structure function. Thus, with this assumption in mind, we will first come up with a different form of the AdS C-metric.

This form, which can be found in (\ref{metric}) below, has two structure functions, each of which is partially factorised in terms of one real root. Moreover, we can use a coordinate freedom in the metric to set these two roots to specific values. The solution is then parameterised in terms of two parameters, which appear as coefficients of the unfactorised part of one of the structure functions. 

The desired coordinate range will then be bound by these two roots. Together with the requirement that the space-time must have Lorentzian signature, this will imply a restricted range for the two parameters. This parameter range can be further divided into a number of subregions, each with a different domain structure. Most of these domains turn out to be triangular in shape, although two are trapezoidal in shape. The box-like domains considered in \cite{Chen:2015vma} also appear as a special case. 

It turns out that one of the trapezoidal domains describes a class of ``slowly accelerating'' spherical black holes in AdS space that has previously been studied in \cite{Podolsky:2002nk,Dias:2002mi,Krtous:2005ej}. Although these black holes are undergoing a constant acceleration, there is no acceleration horizon in the space-time. They are actually static with respect to AdS infinity, with an attached conical singularity providing the necessary tension to counterbalance the cosmological compression of AdS space \cite{Krtous:2005ej}. Moreover, these black holes are not perfectly spherical: they are deformed by the conical singularity that is pulling on it.

On the other hand, the triangular domains collectively describe a class of accelerating or deformed hyperbolic black holes, which is more general than any hitherto considered. Like the above-mentioned deformed spherical black holes, they can be regarded as undergoing an acceleration, although they are effectively static with respect to AdS infinity. In this case, there is no conical singularity in the space-time; since the black holes are connected to AdS infinity, they are able to support themselves and remain static. However, this leads to a distortion of the black-hole horizon, which in some cases can be quite non-trivial.  An extreme case of distortion is the formation of a spherical protrusion out of the otherwise asymptotically hyperbolic horizon.

The organisation of this paper is as follows: We begin in Sec.~\ref{sec2} by explaining how the new form of the AdS C-metric can be derived. In Sec.~\ref{sec3}, we find the complete range of parameters for which this new form describes space-times with Lorentzian signature, and show how it can be divided into a number of subregions. In Sec.~\ref{sec4}, we analyse the domain structure of each subregion, in particular classifying them by the shapes they can take. In Sec.~\ref{sec5}, we focus on one of the possible trapezoidal domains and the triangular domains, and show that they describe deformed spherical and hyperbolic black holes respectively. In Sec.~\ref{sec6}, several special cases of the solution are analysed, while in Sec.~\ref{sec7}, we show how it is related to the traditional forms of the solution. The paper concludes with a summary and discussion of the results.

\section{Derivation of the metric}
\label{sec2}

We begin with the following form of the AdS C-metric proposed in \cite{Chen:2015vma}:
\begin{subequations}\label{Cmetric_alphabeta}
\begin{align}
  \dif s^2&\eq\frac{-\ell^2(a-\alpha)(a-\beta)(b-\alpha)(b-\beta)}{(x-y)^2}\brac{Q(y)\dif t^2-\frac{\dif y^2}{Q(y)}+\frac{\dif x^2}{P(x)}+P(x)\dif\phi^2},
\end{align}
where the structure functions $P(x)$ and $Q(y)$ are cubic polynomials in $x$ and $y$ respectively, given by
\begin{align}
  P(x)&\eq(x-\alpha)(x-\beta)\sbrac{(a+b-\alpha-\beta)(x-a-b)+ab-\alpha\beta},\nonumber\\
  Q(y)&\eq(y-a)(y-b)\sbrac{(a+b-\alpha-\beta)(y-\alpha-\beta)+ab-\alpha\beta}.
\end{align}
\end{subequations}
Here, $\ell$ is the AdS length scale, while $\alpha$, $\beta$, $a$ and $b$ are another four parameters of the solution. These four parameters are deliberately chosen so that the former two are simply roots of $P(x)$, while the latter two are simply roots of $Q(y)$. The third roots of $P(x)$ and $Q(y)$ are then given in terms of them by
\begin{align}
\label{third_roots}
\gamma\equiv a+b+\frac{\alpha\beta-ab}{a+b-\alpha-\beta}\,,\qquad
c\equiv\alpha+\beta+\frac{\alpha\beta-ab}{a+b-\alpha-\beta}\,,
\end{align}
respectively.

As described in \cite{Chen:2015vma}, the metric (\ref{Cmetric_alphabeta}) possesses two continuous symmetries, corresponding to a translation and rescaling of the coordinates $x$ and $y$. It also possesses two discrete symmetries, one corresponding to a reflection of $x$ and $y$, and the other corresponding to the interchanging of any pair of roots. These symmetries can be used to set two of the four independent roots to specific values. In \cite{Chen:2015vma}, we focussed on the case in which $P(x)$ and $Q(y)$ had at least two real roots each. The choice $\alpha=-1$ and $\beta=+1$ was then made on the two roots of $P(x)$, resulting in a slight simplification of the metric (\ref{Cmetric_alphabeta}).

In this paper, we will focus on the more general case in which $P(x)$ and $Q(y)$ have at least one real root each.  Without loss of generality, we take the real root to be $\gamma$ and $c$ respectively. The remaining roots $\alpha$ and $\beta$ of $P(x)$, and $a$ and $b$ of $Q(y)$, may either be real or complex. A real metric is ensured if $\alpha^*=\beta$ and $a^*=b$.

With these assumptions, we can now make use of the above-mentioned symmetries of the metric (\ref{Cmetric_alphabeta}) to make some simplifications to it. The translational symmetry can be used to set $c$ in (\ref{third_roots}) to be zero. This gives
\begin{align}
\gamma=a+b-\alpha-\beta\,,
\end{align}
which we set to be $-1$ using a combination of the reflection and rescaling symmetries. The resulting metric can then be written in terms of just three parameters, say $\alpha$ and $\beta$, in addition to $\ell$. A further simplification can be made by introducing new parameters $\mu$ and $\nu$ by
\begin{align}
\frac{1}{\mu}=\alpha\beta\,,\qquad \frac{\nu}{\mu}=-(\alpha+\beta)\,.
\end{align}
After rescaling $t$ and $\phi$ appropriately, we obtain the metric
\begin{subequations}\label{metric}
\begin{align}
\label{metric1}
 \dif s^2=\frac{\ell^2}{(x-y)^2}\brac{F(y)\dif t^2-\frac{\dif y^2}{F(y)}+\frac{\dif x^2}{G(x)}+G(x)\dif\phi^2},
\end{align}
where the structure functions $F(y)$ and $G(x)$ are given by
\begin{align}
\label{metric2}
 F(y)&=y\sbrac{1+\nu+(\mu+\nu)y+\mu y^2},\nonumber\\
 G(x)&=(1+x)\brac{1+\nu x+\mu x^2}.
\end{align}
\end{subequations}
This is the form of the AdS C-metric that will be used in this paper.

It can be verified that this metric is a solution to the Einstein equation with a cosmological constant $\Lambda=-\frac{3}{\ell^2}$. For future reference, we note that the roots of the structure functions are given by
\begin{subequations}
\begin{align}
 F(y)=0:&\qquad y_0=0,\quad y_\pm=\frac{-(\mu+\nu)\pm\sqrt{(\mu-\nu)^2-4\mu}}{2\mu}\,,\label{F_roots}\\
 G(x)=0:&\qquad x_0=-1,\quad x_\pm=\frac{-\nu\pm\sqrt{\nu^2-4\mu}}{2\mu}\,. \label{G_roots}
\end{align}
\end{subequations}

It can be seen from (\ref{metric1}) that asymptotic infinity lies at $x=y$. On the other hand, from the Kretschmann invariant of the space-time:
\begin{align}
 R_{\mu\nu\rho\sigma}R^{\mu\nu\rho\sigma}\eq\frac{24}{\ell^4}+\frac{12\mu^2(x-y)^6}{\ell^4}\,, \label{Kret}
\end{align}
we see that there are curvature singularities at $x,y\rightarrow\pm\infty$ if $\mu$ is nonzero. In this paper, we will be primarily interested in the domain bounded by the lines $x=x_0$, $y=y_0$ and $x=y$. This gives the general coordinate range
\begin{align}
\label{coord_range}
-1<x<y<0\,,
\end{align}
which will describe a space-time free of curvature singularities and containing an asymptotic infinity. 

The metric (\ref{metric}) clearly has two commuting Killing vectors $\frac{\partial}{\partial t}$ and $\frac{\partial}{\partial\phi}$. It turns out to be also invariant under the transformation
\begin{align}
\label{discrete}
x\rightarrow-y-1\,,\qquad y\rightarrow-x-1\,,\qquad\mu\rightarrow\mu\,,\qquad\nu\rightarrow\mu-\nu\,,
\end{align}
followed by a double-Wick rotation $t\rightarrow i\phi$ and $\phi\rightarrow it$. Note that (\ref{discrete}) has the effect of interchanging the roots:
\begin{align}
\label{discrete1}
x_0\leftrightarrow y_0\,,\qquad x_\pm\leftrightarrow y_\mp\,.
\end{align}
The implications of this discrete symmetry will be discussed below.

\section{Parameter ranges}
\label{sec3}

We now wish to find the full range of parameters $\mu$ and $\nu$ such that (\ref{metric}) describes a space-time with the correct Lorentzian signature ($-$+++), with $\frac{\partial}{\partial t}$ a time-like Killing vector. This means that we require $F<0$ and $G>0$ in the coordinate range (\ref{coord_range}). In this section, we will show how this translates to constraints on the range of $\mu$ and $\nu$.

We first observe that $F$ and $G$ are cubic polynomials of the form
\begin{align}
 f(\xi)\eq\mu\brac{\xi-\xi_1}\brac{\xi-\xi_2}\brac{\xi-\xi_3}.
\end{align}
Assuming that all the roots are real, we can impose the ordering $\xi_1<\xi_2<\xi_3$. Elementary calculus then tells us that, for $\mu<0$, the first derivatives carry the following signs at the roots: $f'(\xi_1)<0$, $f'(\xi_2)>0$ and $f'(\xi_3)<0$, while the second derivatives at the left and right roots satisfy $f''(\xi_1)>0$ and $f''(\xi_3)<0$.\footnote{The second derivative of the middle root can be either positive or negative, depending on whether this root occurs before or after the inflection point of $f$.} For the case $\mu>0$, the inequalities are reversed. We shall denote the roots of $F$ in increasing order as $\{y_1,y_2,y_3\}$, and similarly those of $G$ in increasing order as $\{x_1,x_2,x_3\}$.

\begin{figure}
 \begin{center}
  \begin{subfigure}[b]{0.45\textwidth}
   \centering
   \includegraphics[scale=0.8]{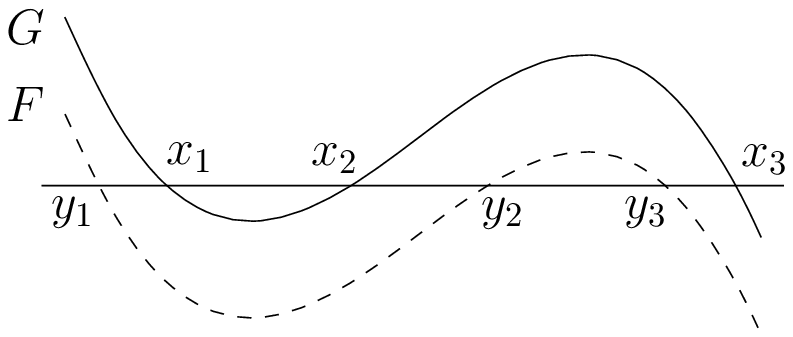}
   \caption{$\mu<0$}
   \label{fig1a}
  \end{subfigure}
  \begin{subfigure}[b]{0.45\textwidth}
   \centering
   \includegraphics[scale=0.8]{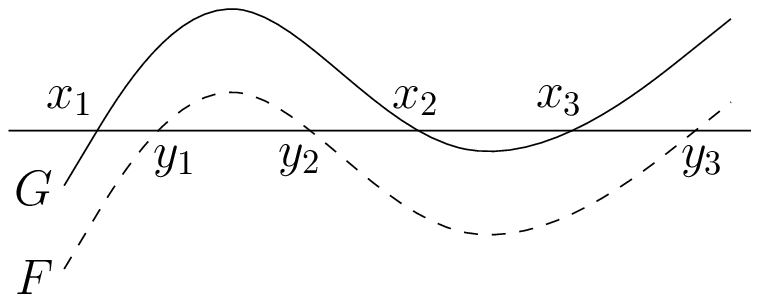}
   \caption{$\mu>0$}
   \label{fig1b}
  \end{subfigure}
 \end{center}
 \caption{The curves of $F(x)$ and $G(x)$ for (a) $\mu<0$, and (b) $\mu>0$.}
 \label{fig1}
\end{figure}

Since $F$ and $G$ satisfy the identity 
\begin{align}
G(x)-F(x)=1\,,
\end{align}
they share the same polynomial coefficients, except for the constant term which differs by 1. Thus when sketched on a common axis, the curves $F$ and $G$ have the same profile but are shifted vertically from each other by a unit distance, as shown in Fig.~\ref{fig1} separately for the cases $\mu\lessgtr 0$.

From this, we conclude that the order of the roots for $\mu<0$ is
\begin{align}
 y_1<x_1<x_2<y_2<y_3<x_3\,, \label{CB_root_order2}
\end{align}
while for $\mu>0$ it is
\begin{align}
 x_1<y_1<y_2<x_2<x_3<y_3\,. \label{CB_root_order1}
\end{align}

At this stage, we remark that the above results can also be extended to the case of complex roots. For example, in the case $\mu<0$, it is possible for the graph of $G$ in Fig.~\ref{fig1a} to not intersect the axis at $x_1$ and $x_2$, corresponding to these two roots of $G$ becoming complex. It is also possible for the graph of $F$ to not intersect the axis at $y_2$ and $y_3$, corresponding to these two roots of $F$ becoming complex. In such a situation, we simply remove $x_1$ and $x_2$, or $y_2$ and $y_3$, from the ordering (\ref{CB_root_order2}). Similar remarks apply in the case $\mu>0$, with respect to the roots $y_1$ and $y_2$, or $x_2$ and $x_3$.

With these considerations in mind, we are now ready to identify the roots of $F$, namely $\{y_1,y_2,y_3\}$, with the possible choices $\{y_0,y_+,y_-\}$ given by (\ref{F_roots}); and similarly identify the roots of $G$, namely $\{x_1,x_2,x_3\}$, with the possible choices $\{x_0,x_+,x_-\}$ given by (\ref{G_roots}). In particular, the positions of the roots $y_0=0$ and $x_0=-1$ relative to the other roots will then give constraints on $\mu$ and $\nu$ via the evaluation of $F'(0)$, $F''(0)$, $G'(-1)$ and $G''(-1)$, and subjecting them to the above-stated inequalities. 

\begin{figure}
 \begin{center}
  \includegraphics[scale=0.6]{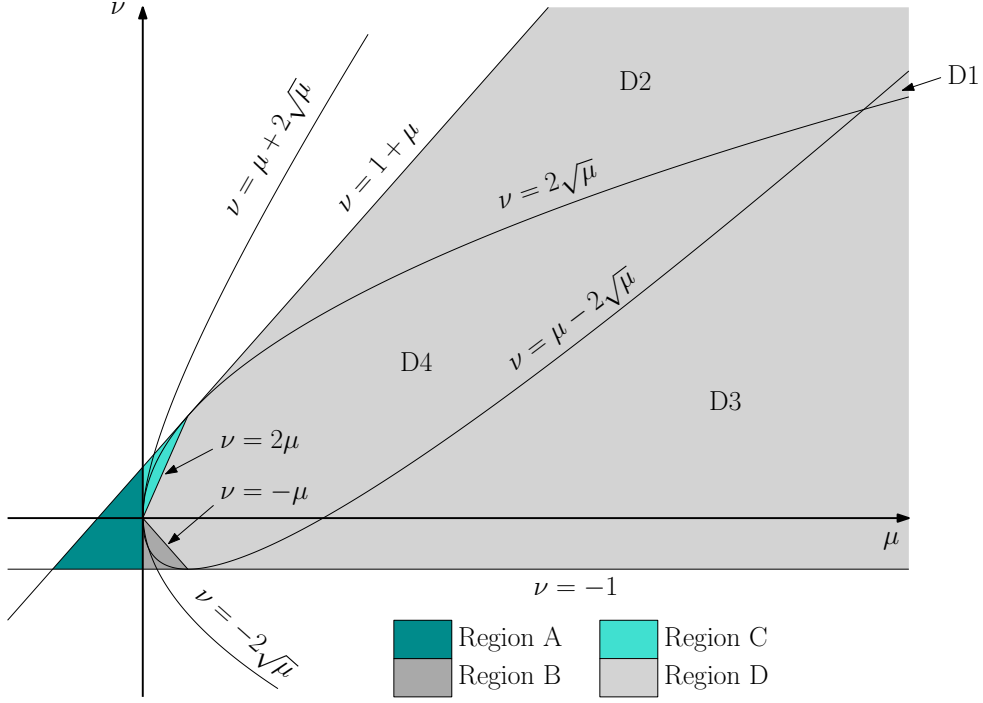}
  \caption{The parameter range as a plot of $\nu$ versus $\mu$. The four shaded regions A--D are those which describe a Lorentzian space-time in the coordinate range $-1<x<y<0$.}
  \label{fig2}
 \end{center}
\end{figure}

We begin with the case $\mu<0$, with the curves of $F$ and $G$ as depicted in Fig.~\ref{fig1a}. If we wish to have a space-time that is Lorentzian in the range (\ref{coord_range}), the only possibility is to set $y_2=0$ and $x_2=-1$. In other words, $0$ and $-1$ are the middle roots of $F$ and $G$ respectively. The conditions for this to occur are
\begin{align}
\label{F1G1}
F'(0)=1+\nu>0\,,\qquad G'(-1)=1+\mu-\nu>0\,.
\end{align}
The resulting range of parameters is therefore $\mu<0$, $\nu>-1$ and $\nu<1+\mu$, and is labelled by Region A in the $\mu$--$\nu$ plot of Fig.~\ref{fig2}. In this case, note that the ordering of the roots (\ref{CB_root_order2}) translates to
\begin{align}
\label{order_A}
 y_+<x_+<-1<0<y_-<x_-\,.
\end{align}

We now turn to the case $\mu>0$, with the curves of $F$ and $G$ as depicted in Fig.~\ref{fig1b}. In this case, if we wish to have a Lorentzian space-time in the range (\ref{coord_range}), $y=0$ and $x=-1$ {\it cannot\/} be the middle roots. This actually gives the same conditions as in (\ref{F1G1}), since the sign of $\mu$ is opposite to that of the previous case. However, there are now additional conditions coming from the signs of $F''(0)$ and $G''(-1)$, depending on whether $y_0=0$ and $x_0=-1$ are the left or right roots. There are three separate cases to consider:
\begin{itemize}
 \item $y_0=0$ and $x_0=-1$ are both left roots: The additional conditions for this to occur are
\begin{align}
F''(0)=2(\mu+\nu)<0\,,\qquad G''(-1)=2(\nu-2\mu)<0\,.
\end{align}
The resulting range of parameters is therefore $\mu>0$, $\nu>-1$ and $\nu<-\mu$, and is labelled by Region B in Fig.~\ref{fig2}. In this case, note that the ordering of the roots (\ref{CB_root_order1}) translates to
\begin{align}
\label{order_B}
 -1<0<y_-<x_-<x_+<y_+\,.
\end{align}

 \item $y_0=0$ and $x_0=-1$ are both right roots: The additional conditions for this to occur are
\begin{align}
F''(0)=2(\mu+\nu)>0\,,\qquad G''(-1)=2(\nu-2\mu)>0\,.
\end{align}
The resulting range of parameters is therefore $\mu>0$, $\nu<1+\mu$ and $\nu>2\mu$, and is labelled by Region C in Fig.~\ref{fig2}. In this case, note that the ordering of the roots (\ref{CB_root_order1}) translates to
\begin{align}
\label{order_C}
 x_-<y_-<y_+<x_+<-1<0\,.
\end{align}

 \item $y_0=0$ is a right root and $x_0=-1$ a left root: The additional conditions for this to occur are
\begin{align}
F''(0)=2(\mu+\nu)>0\,,\qquad G''(-1)=2(\nu-2\mu)<0\,.
\end{align}
The resulting range of parameters is therefore $\nu>-1$, $\nu<1+\mu$, $\nu>-\mu$ and $\nu<2\mu$, and is labelled by Region D in Fig.~\ref{fig2}. In this case, note that the ordering of the roots (\ref{CB_root_order1}) translates to
\begin{align}
\label{order_D}
 -1<y_-<y_+<x_-<x_+<0\,.
\end{align}
\end{itemize}

We conclude this section by mentioning the effect of the transformation (\ref{discrete}) on the various regions of Fig.~\ref{fig2}. Note that it has the effect of interchanging the inequalities:
\begin{align}
\nu>-1\quad&\leftrightarrow\quad\nu<1+\mu\,,\cr
\nu\lessgtr-\mu\quad&\leftrightarrow\quad\nu\gtrless2\mu\,.
\end{align}
It follows that this transformation maps Region B to Region C, and {\it vice versa\/}. On the other hand, it maps Region A to itself, and similarly for Region D.

\section{Domain structure}
\label{sec4}

In the previous section, we found that in order for (\ref{metric}) to describe a Lorentzian space-time in the coordinate range (\ref{coord_range}), the parameters $\mu$ and $\nu$ must lie in one of the four shaded regions A--D in Fig.~\ref{fig2}. We now turn to an analysis of the domain structure for each region in turn. It will be verified that triangular domains exist in each region, although as we shall see, other types of domains also occur in Region D.

\begin{figure}
\begin{center}
  \includegraphics[scale=0.45]{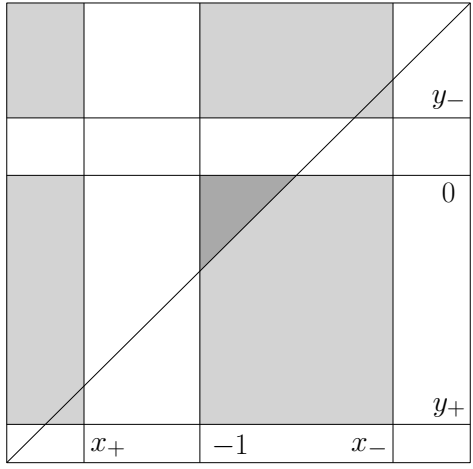}
\end{center}
\caption{The domain structure for Region A. The $x$-axis is in the horizontal direction, while the $y$-axis is in the vertical direction. Shaded areas correspond to coordinate ranges with Lorentzian signature, with the darker shade representing the coordinate range of particular interest $-1<x<y<0$. The diagonal line corresponds to $x=y$, while the edges of the plot correspond to $x,y=\pm\infty$.}
\label{figA}
\end{figure}

\begin{figure}[t]
\begin{center}
 \begin{subfigure}[b]{0.4\textwidth}
  \centering
  \includegraphics[scale=0.45]{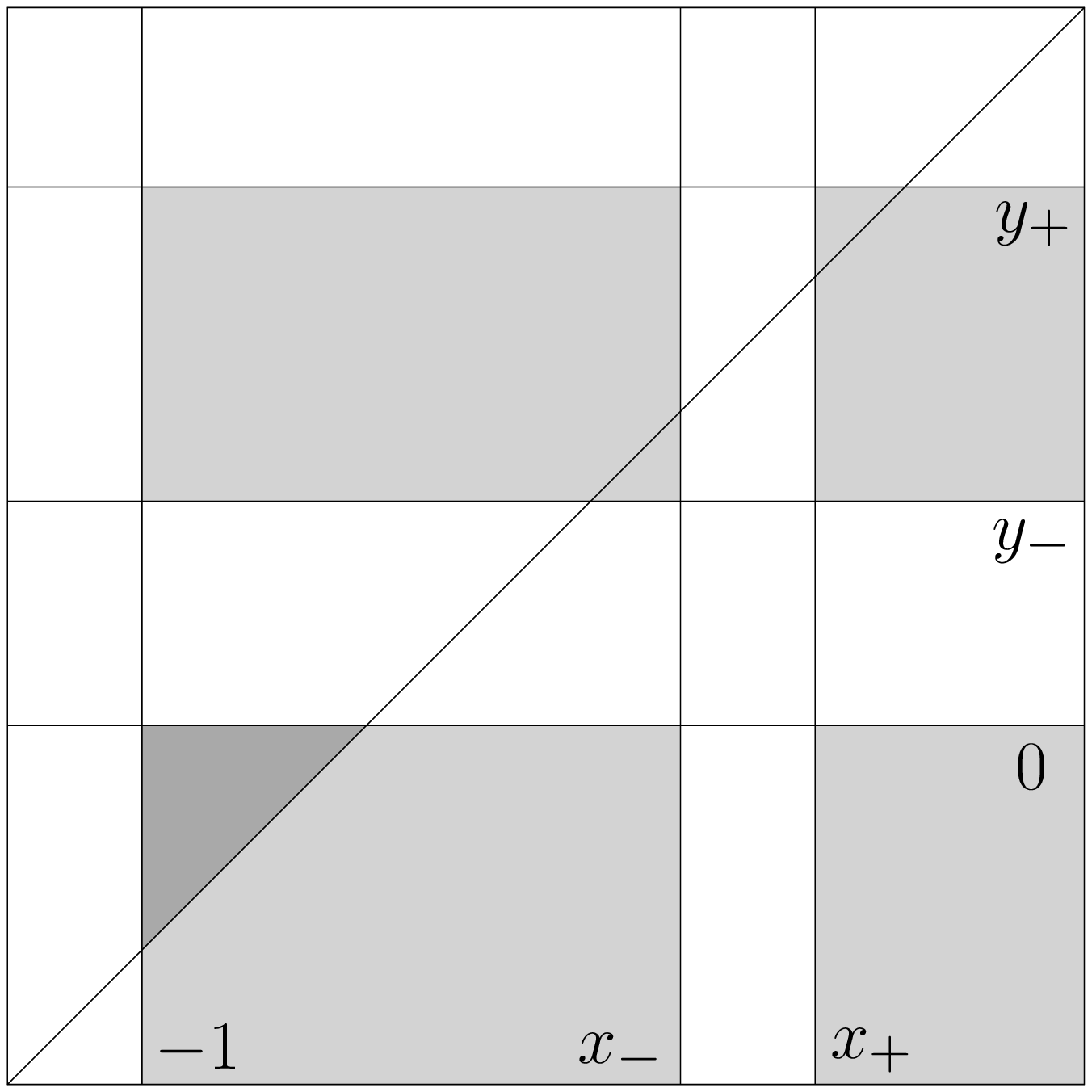}
  \caption{Region B1}
  \label{figB1}
 \end{subfigure}\vspace{6pt}
~~~ \begin{subfigure}[b]{0.4\textwidth}
  \centering
  \includegraphics[scale=0.45]{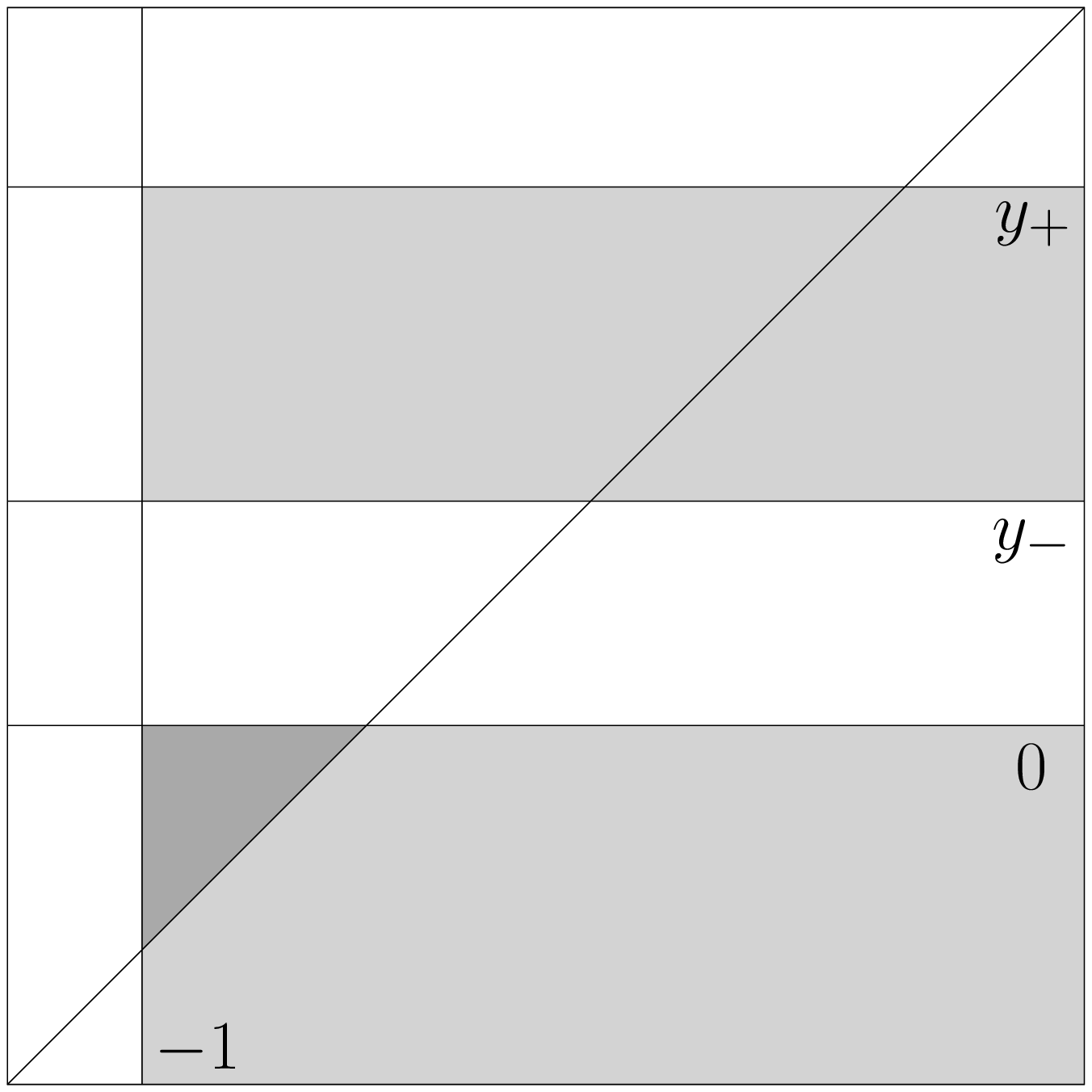}
  \caption{Region B2}
  \label{figB2}
 \end{subfigure}
 \begin{subfigure}[b]{0.4\textwidth}
  \centering
  \includegraphics[scale=0.45]{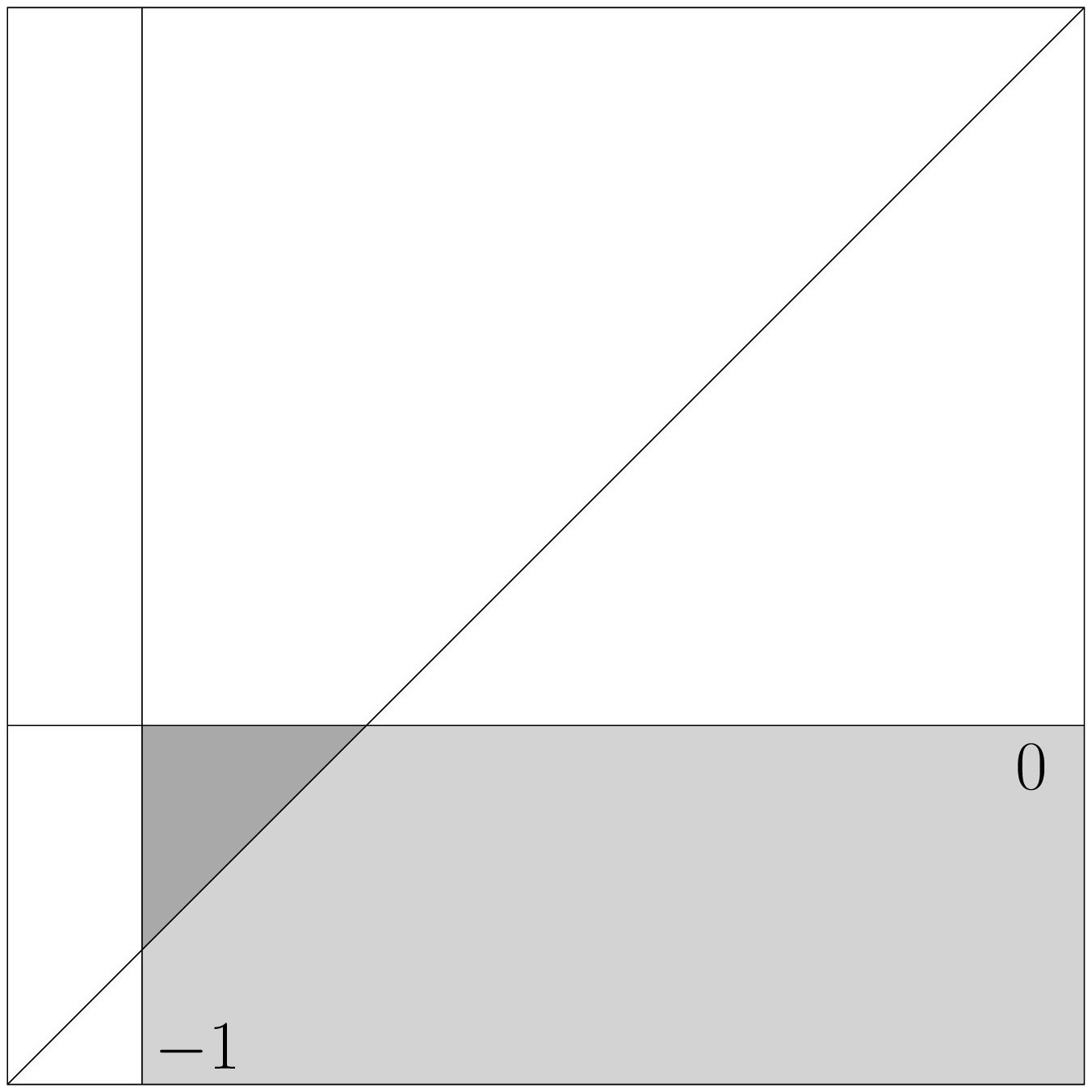}
  \caption{Region B3}
  \label{figB3}
 \end{subfigure}
\end{center}
\caption{The domain structure for the various parts of Region B.}
\label{figB}
\end{figure}

\begin{figure}[t]
\begin{center}
 \begin{subfigure}[b]{0.4\textwidth}
  \centering
  \includegraphics[scale=0.45]{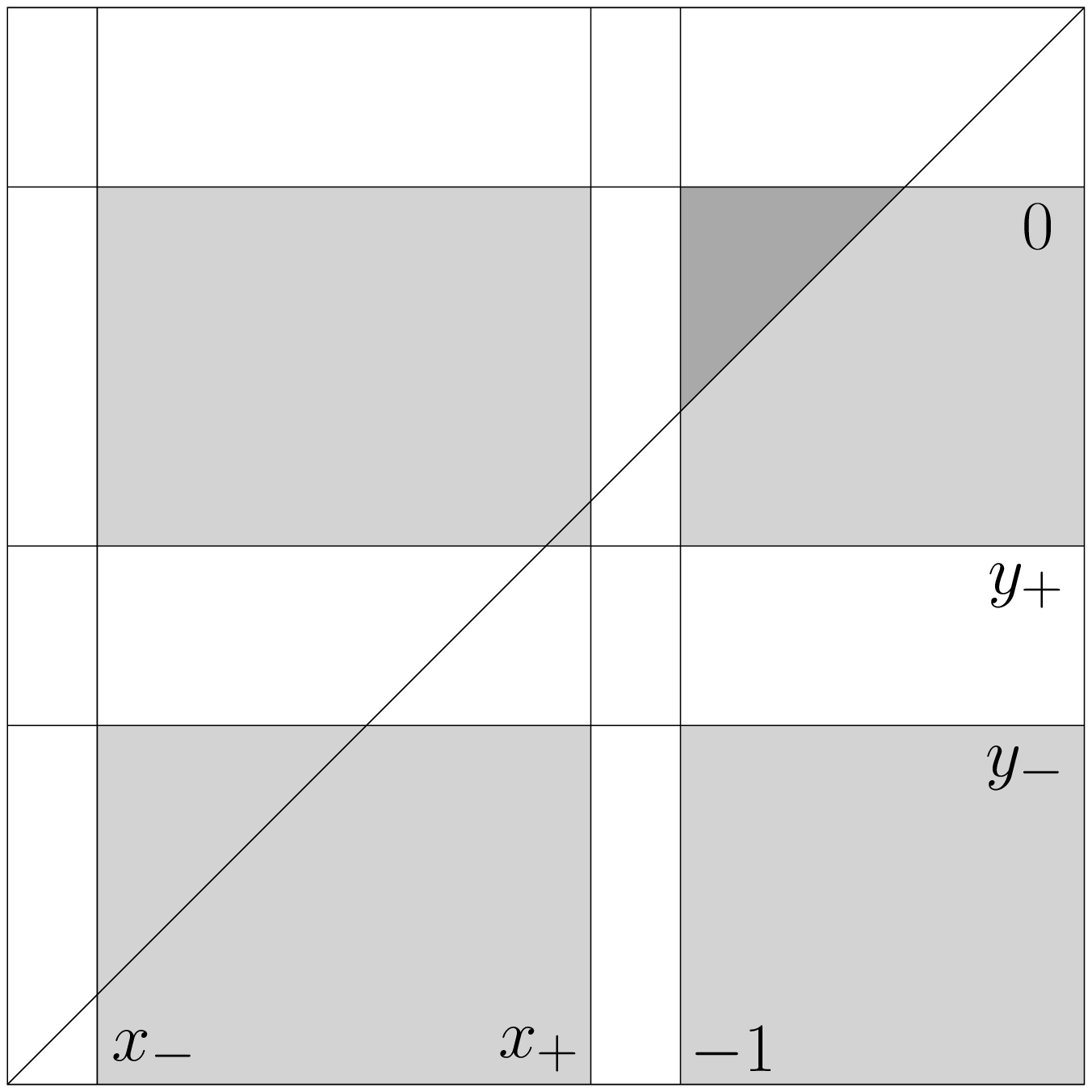}
  \caption{Region C1}
  \label{figC1}
 \end{subfigure}\vspace{6pt}
~~~ \begin{subfigure}[b]{0.4\textwidth}
  \centering
  \includegraphics[scale=0.45]{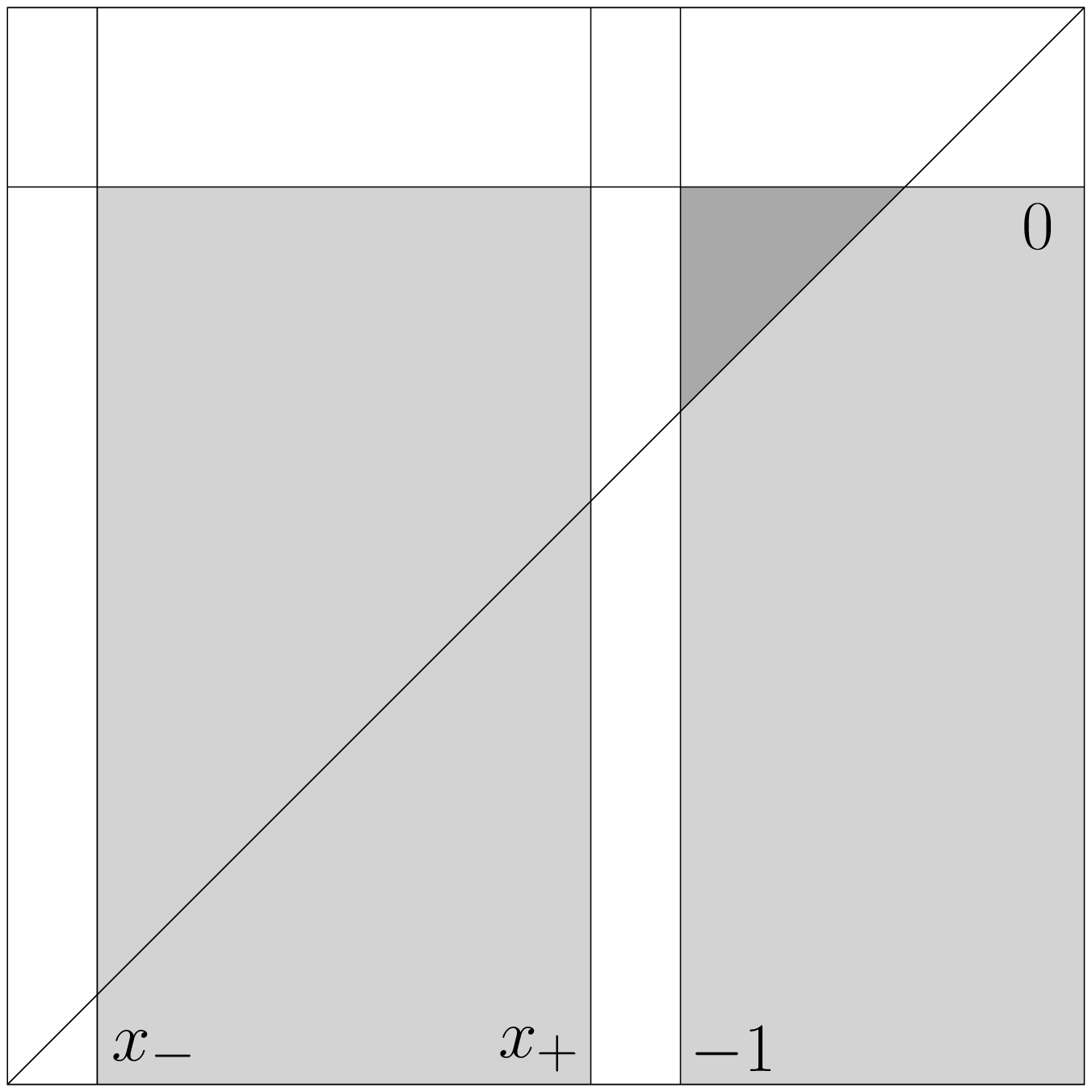}
  \caption{Region C2}
  \label{figC2}
 \end{subfigure}
 \begin{subfigure}[b]{0.4\textwidth}
  \centering
  \includegraphics[scale=0.45]{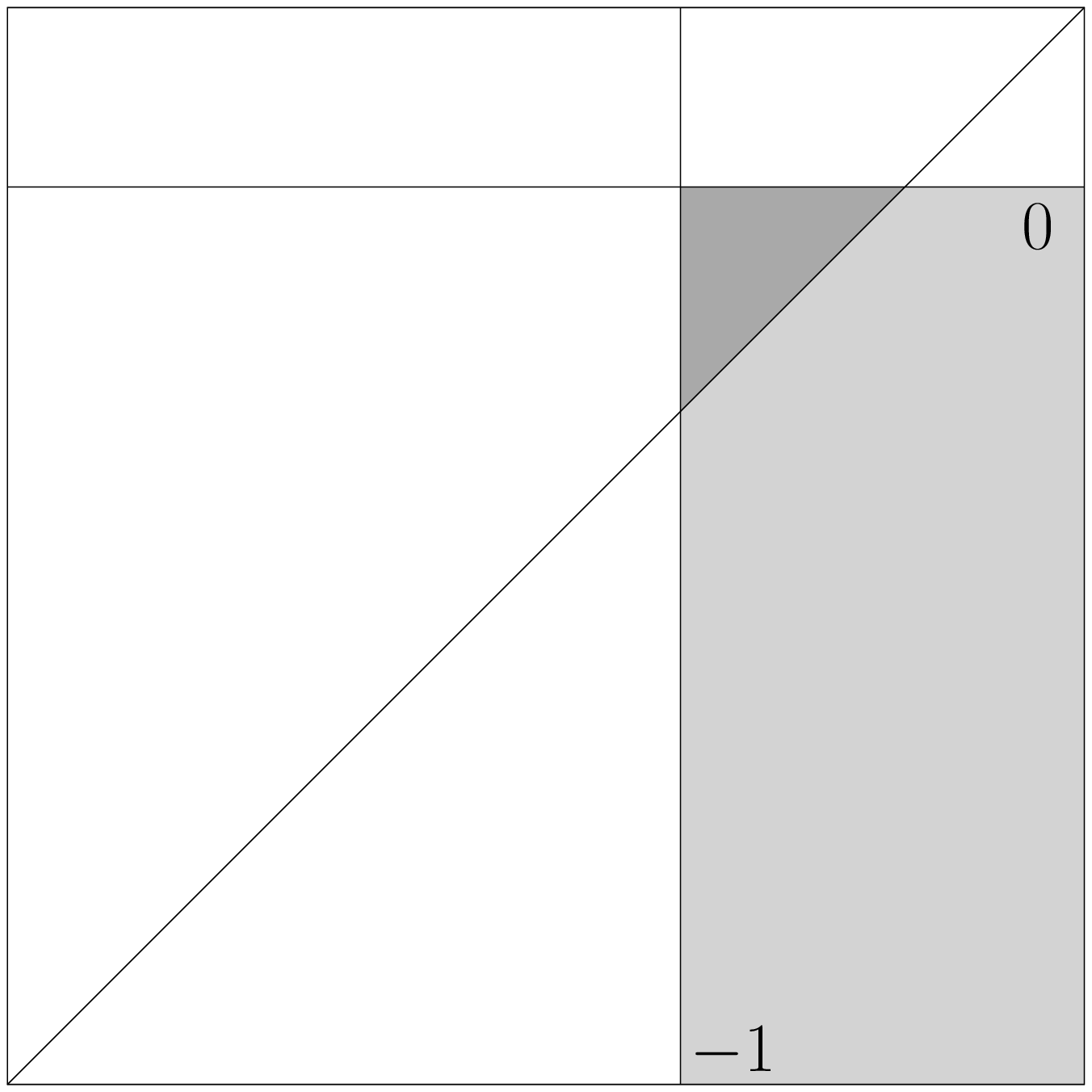}
  \caption{Region C3}
  \label{figC3}
 \end{subfigure}
\end{center}
\caption{The domain structure for the various parts of Region C.}
\label{figC}
\end{figure}

\begin{figure}[t]
\begin{center}
 \begin{subfigure}[b]{0.4\textwidth}
  \centering
  \includegraphics[scale=0.45]{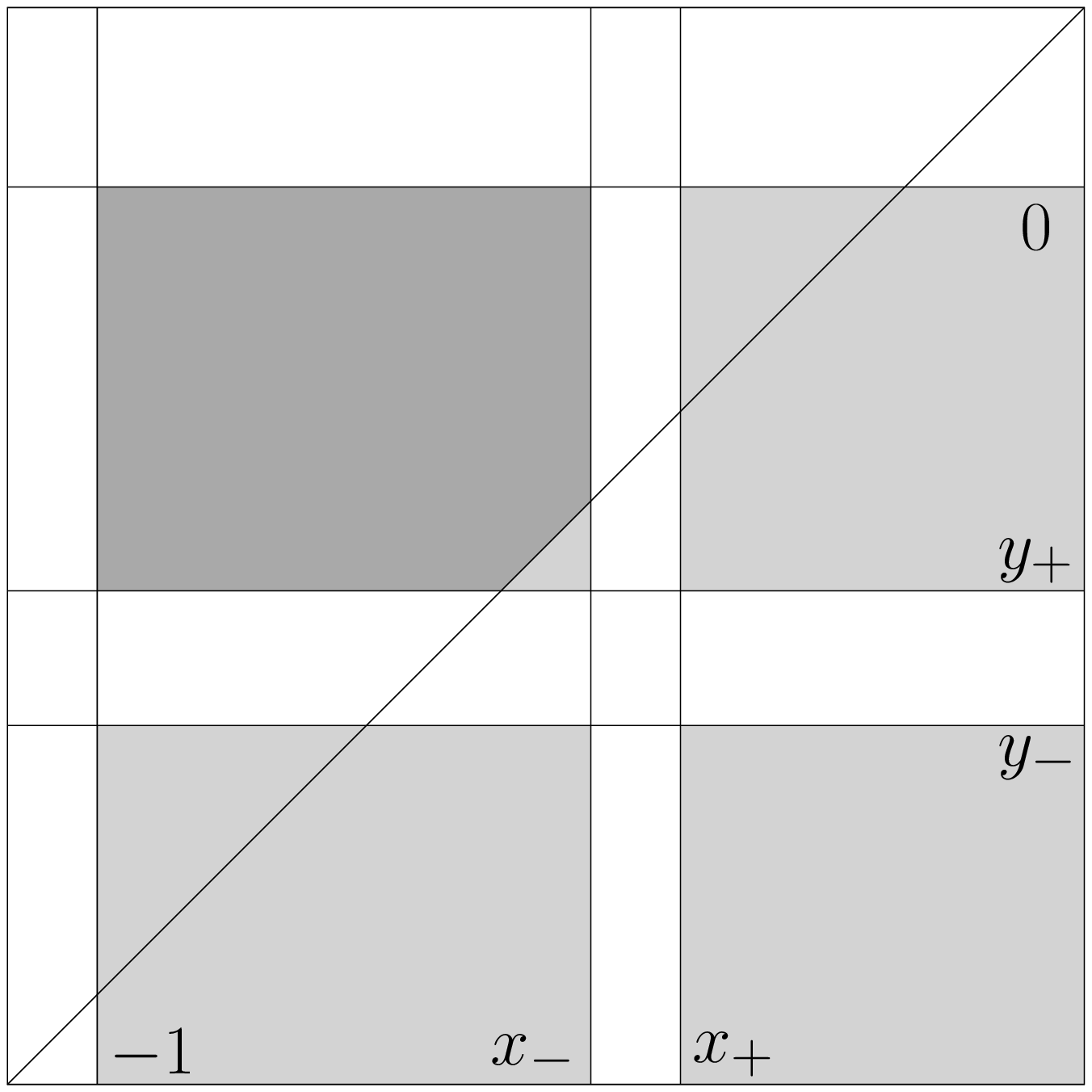}
  \caption{Region D1}
  \label{figD1}
 \end{subfigure}\vspace{6pt}
~~~ \begin{subfigure}[b]{0.4\textwidth}
  \centering
  \includegraphics[scale=0.45]{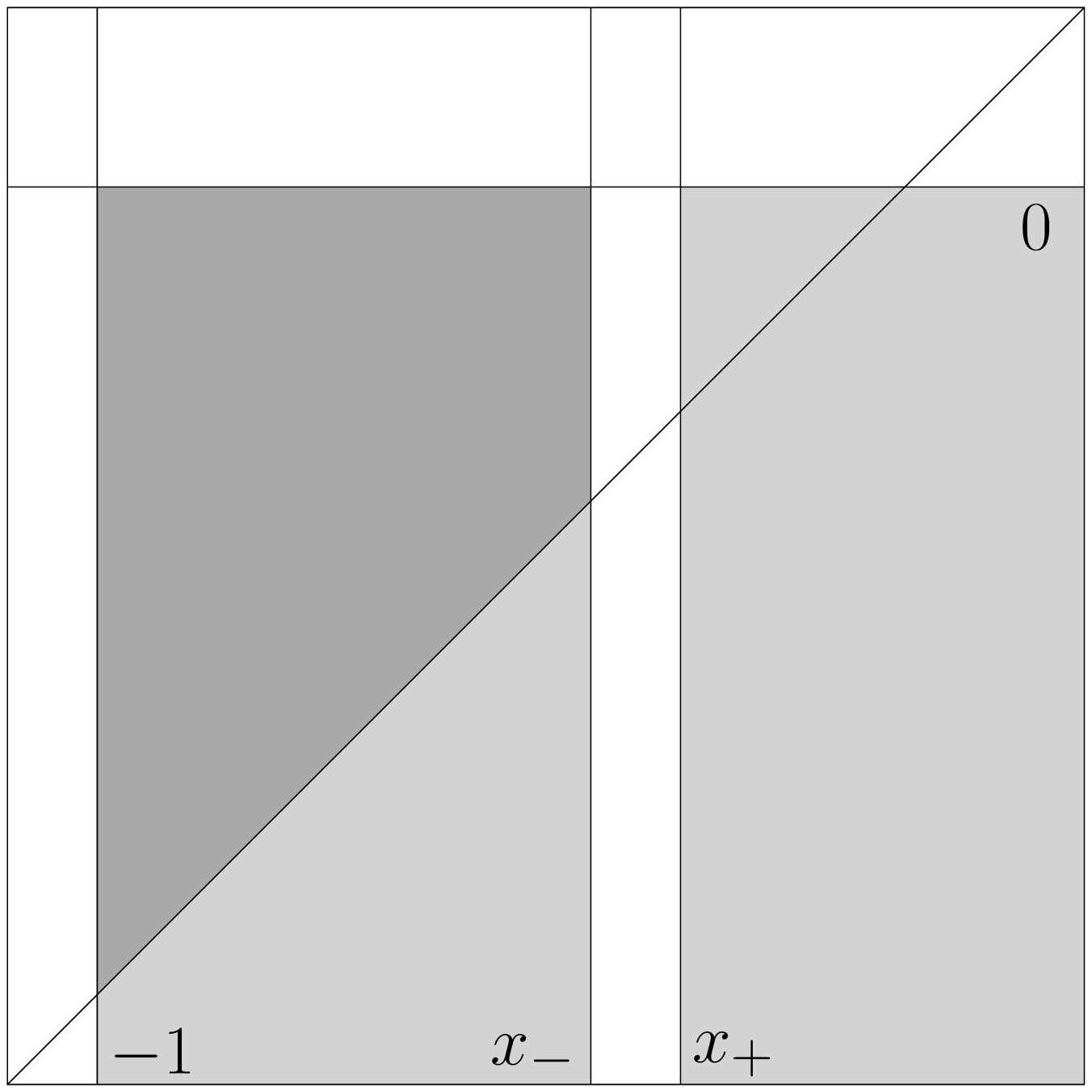}
  \caption{Region D2}
  \label{figD2}
 \end{subfigure}
 \begin{subfigure}[b]{0.4\textwidth}
  \centering
  \includegraphics[scale=0.45]{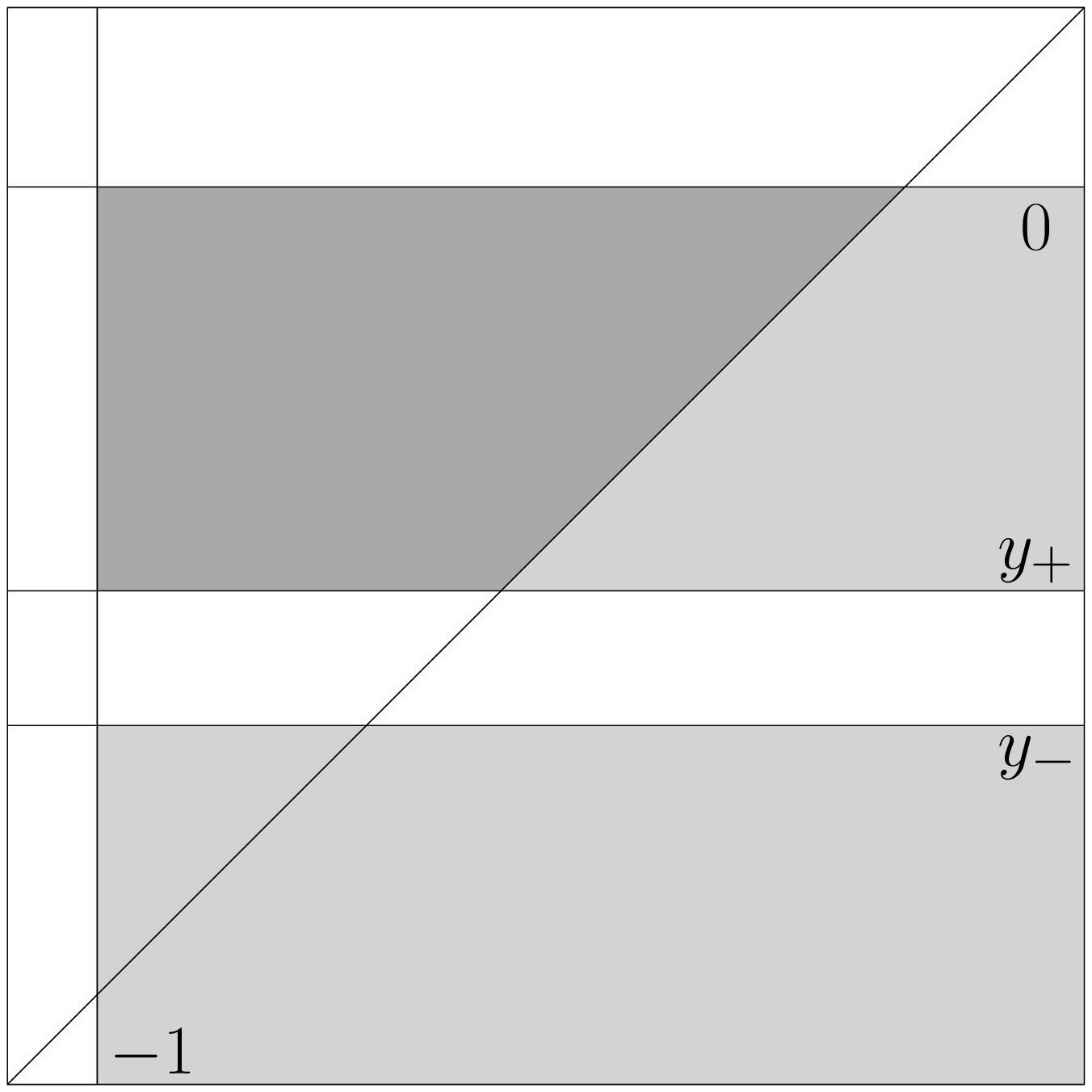}
  \caption{Region D3}
  \label{figD3}
 \end{subfigure}\hspace{12pt}
 \begin{subfigure}[b]{0.4\textwidth}
  \centering
  \includegraphics[scale=0.45]{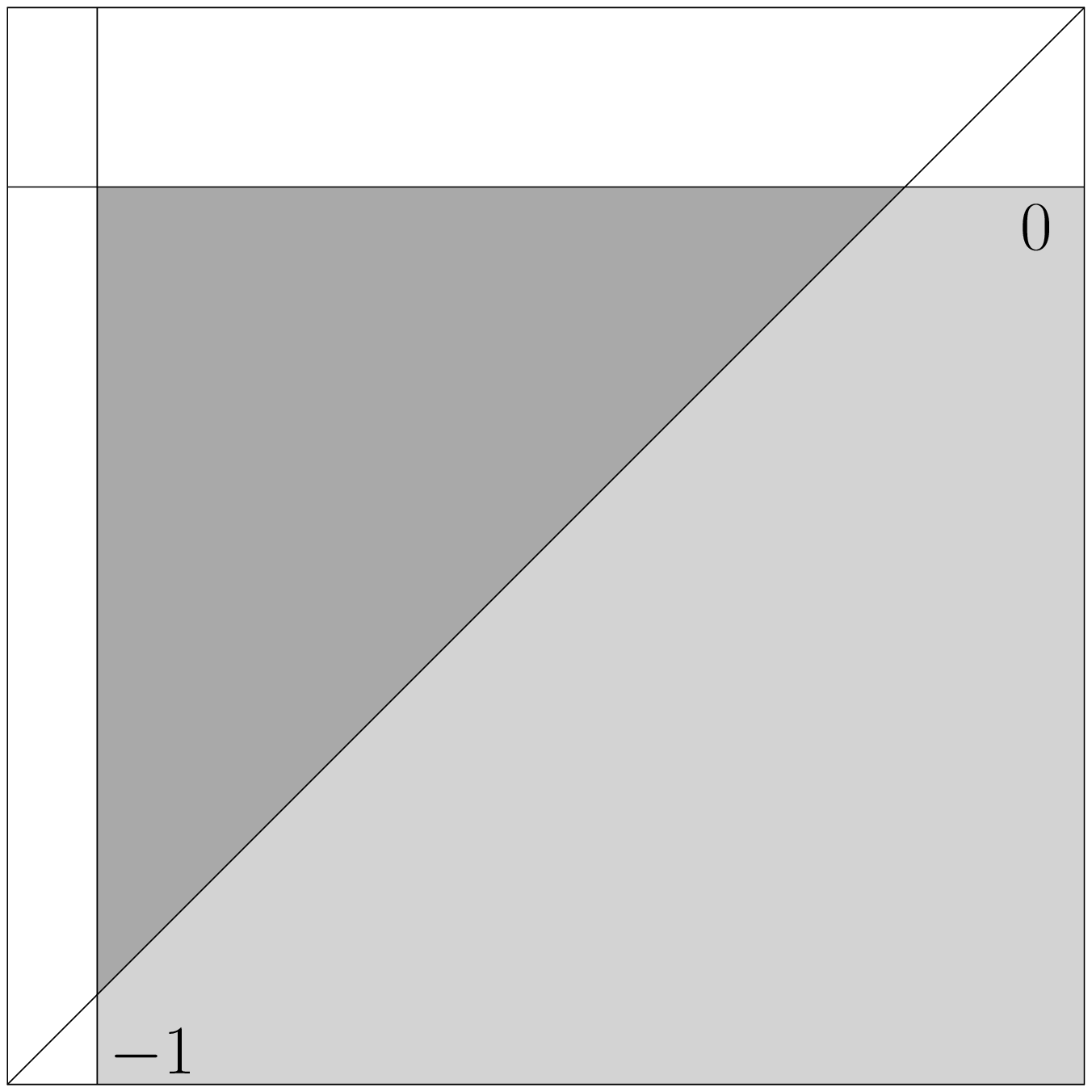}
  \caption{Region D4}
  \label{figD4}
 \end{subfigure}
\end{center}
\caption{The domain structure for the various parts of Region D.}
\label{figD}
\end{figure}

In Region A, the ordering of the roots is given by (\ref{order_A}). In this case, it can be checked that the roots $x_\pm$ and $y_\pm$ are always real. This leads to the domain structure depicted in Fig.~\ref{figA}.\footnote{We mention that diagrams similar to Figs.~\ref{figA}--\ref{figD} have appeared in \cite{Hubeny:2009kz}. In that paper, however, $x-y$ instead of $y$ is plotted as a function of $x$.} The Lorentzian space-time described by the coordinate range (\ref{coord_range}) is then represented by the darker-shaded triangle in this figure. 

In Region B, the ordering of the roots is given by (\ref{order_B}). Unlike the previous case, it is possible for the roots $x_\pm$ and $y_\pm$ to become complex. It can be checked that the roots $x_\pm$ will become complex when $\nu>-2\sqrt\mu$, while the roots $y_\pm$ will become complex when $\nu>\mu-2\sqrt\mu$. This leads to a partitioning of Region B into the three subregions:
  \begin{equation*}
    \begin{array}{rcc}
     \mbox{B1:\quad} & x_\pm\mbox{ and }y_\pm\mbox{ real, } &\nu<-2\sqrt{\mu}\,;\\
     \mbox{B2:\quad}& x_\pm\mbox{ complex}, y_\pm\mbox{ real, } & -2\sqrt{\mu}<\nu<\mu-2\sqrt{\mu}\,;\\
     \mbox{B3:\quad}& x_\pm\mbox{ and }y_\pm\mbox{ complex, } & \nu>\mu-2\sqrt{\mu}\,.
    \end{array}
  \end{equation*}
In Fig.~\ref{fig2}, note that Region B is indeed cut by the curves $\nu=-2\sqrt\mu$ and $\nu=\mu-2\sqrt\mu$. The subregions B1, B2 and B3 are then the left, middle and right parts of this region. 

The domain structure for B1 is depicted in Fig.~\ref{figB1}. The Lorentzian space-time described by the coordinate range (\ref{coord_range}) is again represented by the darker-shaded triangle in this figure. The roots $x_\pm$ in Fig.~\ref{figB1} will coalesce into a single degenerate root if $\nu=-2\sqrt\mu$, and will disappear altogether for $\nu>-2\sqrt{\mu}$. This results in the domain structure depicted for B2 in Fig.~\ref{figB2}. If $\nu$ is further increased to $\nu=\mu-2\sqrt{\mu}$, the roots $y_\pm$ will also coalesce, and will disappear for $\nu>\mu-2\sqrt{\mu}$. This results in the domain structure for B3 depicted in Fig.~\ref{figB3}.

In Region C, the ordering of the roots is given by (\ref{order_C}). As in the previous case, it is possible for the roots $x_\pm$ and $y_\pm$ to become complex. In this case, the roots $x_\pm$ will become complex when $\nu<2\sqrt\mu$, while the roots $y_\pm$ will become complex when $\nu<\mu+2\sqrt\mu$. This leads to a partitioning of Region C into the three subregions:
  \begin{equation*}
    \begin{array}{rcc}
     \mbox{C1:\quad} & x_\pm\mbox{ and }y_\pm\mbox{ real, } &\nu>\mu+2\sqrt{\mu}\,;\\
     \mbox{C2:\quad}& x_\pm\mbox{ real}, y_\pm\mbox{ complex, } & 2\sqrt{\mu}<\nu<\mu+2\sqrt{\mu}\,;\\
     \mbox{C3:\quad}& x_\pm\mbox{ and }y_\pm\mbox{ complex, } & \nu<2\sqrt{\mu}\,.
    \end{array}
  \end{equation*}
In Fig.~\ref{fig2}, note that Region C is indeed cut by the curves $\nu=2\sqrt\mu$ and $\nu=\mu+2\sqrt\mu$. The subregions C1, C2 and C3 are then the left, middle and right parts of this region. 

The domain structure for C1 is depicted in Fig.~\ref{figC1}. The Lorentzian space-time described by the coordinate range (\ref{coord_range}) is again represented by the darker-shaded triangle in this figure. The roots $y_\pm$ in Fig.~\ref{figC1} will coalesce into a single degenerate root if $\nu=\mu+2\sqrt\mu$, and will disappear altogether for $\nu<\mu+2\sqrt\mu$. This results in the domain structure depicted for C2 in Fig.~\ref{figC2}. If $\nu$ is further decreased to $\nu=2\sqrt{\mu}$, the roots $x_\pm$ will also coalesce, and will disappear for $\nu<2\sqrt{\mu}$. This results in the domain structure for C3 depicted in Fig.~\ref{figC3}.

In Region D, the ordering of the roots is given by (\ref{order_D}). As in the previous two cases, it is possible for the roots $x_\pm$ and $y_\pm$ to become complex. In this case, the roots $x_\pm$ will become complex when $\nu<2\sqrt\mu$, while the roots $y_\pm$ will become complex when $\nu>\mu-2\sqrt\mu$. This leads to a partitioning of Region D into the four subregions:
  \begin{equation*}
    \begin{array}{rcc}
     \mbox{D1:\quad} & x_\pm\mbox{ and }y_\pm\mbox{ real, } & 2\sqrt{\mu}<\nu<\mu-2\sqrt{\mu}\,;\\
     \mbox{D2:\quad}& x_\pm\mbox{ real}, y_\pm\mbox{ complex, } & \nu>2\sqrt{\mu}\mbox{ and }\nu>\mu-2\sqrt{\mu}\,;\\
     \mbox{D3:\quad}& x_\pm\mbox{ complex}, y_\pm\mbox{ real, } & \nu<2\sqrt{\mu}\mbox{ and }\nu<\mu-2\sqrt{\mu}\,;\\
     \mbox{D4:\quad}& x_\pm\mbox{ and }y_\pm\mbox{ complex, } & \mu-2\sqrt{\mu}<\nu<2\sqrt{\mu}\,.
    \end{array}
  \end{equation*}
In Fig.~\ref{fig2}, note that Region D is indeed cut by the curves $\nu=2\sqrt\mu$ and $\nu=\mu-2\sqrt\mu$. The subregions D1, D2, D3 and D4 are then the parts of this region as indicated in Fig.~\ref{fig2}. 

The domain structure for D1 is depicted in Fig.~\ref{figD1}. In this case, because the roots $x_\pm$ and $y_\pm$ lie between $-1$ and 0, the coordinate range (\ref{coord_range}) describes a Lorentzian space-time only if $-1<x<x_-$ and $y_+<y<0$. The domain of interest is then a five-sided ``box'' bounded by the lines $x=-1$, $x=x_-$ and $y=0$, $y=y_+$, as well as the line $x=y$. The roots $y_\pm$ in Fig.~\ref{figD1} will coalesce and become complex if $\nu>\mu-2\sqrt\mu$, and similarly for the roots $x_\pm$ if $\nu<2\sqrt{\mu}$. This results in the domain structures for D2, D3 and D4, depicted in Figs.~\ref{figD2}, \ref{figD3} and \ref{figD4}, respectively. Note that the domains of interest for D2 and D3 are trapezoids, while that for D4 is a triangle.

The effect of the transformation (\ref{discrete}) on the various regions of Fig.~\ref{fig2} can also be seen from their domain structures in Figs.~\ref{figA}--\ref{figD}. Note that the transformation of $x$ and $y$ in (\ref{discrete}) effectively ``flips'' the domain structures about the diagonal line joining the upper-left and lower-right corners. The exact mapping between the roots $\{x_0,x_\pm\}$ and $\{y_0,y_\pm\}$ is given by (\ref{discrete1}). In this sense, we see that the domain structure of Region A is mapped to itself. We also see that the domain structures of Regions B1, B2 and B3 are mapped to those of Regions C1, C2 and C3 respectively, and {\it vice versa\/}. Turning to Region D, we see that the domain structure of D2 is mapped to that of D3, and {\it vice versa\/}. On the other hand, the domain structure of D1 is mapped to itself, and similarly for D4.

To summarise, we have seen that triangular domains arise from Regions A, B, C and D4. The differences between them lies in the presence or absence of the roots $x_\pm$ and $y_\pm$, and their locations if they are present. In Regions A, B1 and C1, the roots $x_\pm$ and $y_\pm$ are all present, but their locations relative to the roots $x_0=-1$ and $y_0=0$ are different in each case. In Region B2, only the roots $y_\pm$ are present, while in Region C2, only the roots $x_\pm$ are present. In Regions B3, C3 and D4, the roots $x_\pm$ and $y_\pm$ are all absent. Together, these regions form all the possible different domain structures with triangular domains. It is worth noting that these regions form a single connected region in the plot of Fig.~\ref{fig2}.

We have also seen that trapezoidal domains arise in Regions D2 and D3, while a box-like domain arises in Region D1. The physics of the space-times described by each of these types of domains is different. In \cite{Chen:2015vma}, the focus was on the box-like domain, and it was shown that it describes a class of black holes in AdS space undergoing a constant acceleration, in the presence of an acceleration horizon. In Sec.~\ref{sec5.1}, we will discuss the physics of the space-times described by the trapezoidal domain of Region D2. It will be shown that it describes the class of deformed spherical black holes that was studied in \cite{Podolsky:2002nk,Dias:2002mi,Krtous:2005ej}. Unlike those described by Region D1, these black holes do not have acceleration horizons associated to them.

The main focus of this paper however, will be on the triangular domains arising from Regions A, B, C and D4.  In Sec.~\ref{sec5.2}, we will discuss the physics of the space-times described by such domains. It will be shown that they describe deformed hyperbolic black holes. They are in fact hyperbolic analogues of the deformed spherical black holes described by Region D2.

\section{Physical interpretation}
\label{sec5}

\subsection{Deformed spherical black holes}
\label{sec5.1}

We begin with the trapezoidal domain of Region D2. As explained in \cite{Chen:2015vma}, the shape of the domain contains much useful physical information about the space-time. Recall that the left and right edges of the trapezoid are points at which the metric coefficient $g_{\phi\phi}$ vanishes; they represent the two symmetry axes of the space-time. The upper edge of the trapezoid are points at which $g_{tt}$ vanishes; it represents a horizon of the space-time. On the other hand, the lower edge of the trapezoid represents asymptotic infinity. It follows that we have a finite horizon that separates the two asymptotic axes. This horizon has $S^2$ topology, so we conclude that the space-time contains a spherical black hole.

More detailed information on the geometric properties of this space-time can be obtained by calculating its so-called rod structure (see \cite{Chen:2010zu} and references therein). In this case, the rod structure consists of the following three rods: 
\begin{itemize}
 \item Rod 1: a semi-infinite space-like rod located at ($x=-1,-1<y\leq 0$), with direction $\ell_1=\frac{1}{\kappa_{\mathrm{1}}}\frac{\partial}{\partial\phi}$, where
  \begin{align}
    \label{kappa1}
   \kappa_{\mathrm{1}}=\half\brac{\mu-\nu+1}.
  \end{align}
  \item Rod 2: a finite time-like rod located at ($-1\leq x\leq x_-,y=0$), with direction $\ell_2=\frac{1}{\kappa_2}\frac{\partial}{\partial t}$, where
    \begin{align}
     \kappa_2=\half\brac{\nu+1}.
    \end{align}
\item Rod 3: a semi-infinite space-like rod located at ($x=x_-,x_-<y\leq 0$), with direction $\ell_3=\frac{1}{\kappa_{\mathrm{3}}}\frac{\partial}{\partial\phi}$, where
  \begin{align}
    \label{kappa3}
   \kappa_{\mathrm{3}}=\frac{\sqrt{\nu^2-4\mu}}{4\mu}\left(2\mu-\nu-\sqrt{\nu^2-4\mu}\right).
  \end{align}
\end{itemize}
It can be seen that Rods 1 and 3 are the two asymptotic axes, while Rod 2 is the black-hole horizon. The directions of the rods encode information about the axes and horizon. In particular, the normalisation factors (\ref{kappa1}) and (\ref{kappa3}) encode the natural periodicity of the $\phi$ coordinate around the two axes. To avoid a conical singularity along Rod 1 or Rod 3, the identification
\begin{align}
(t,\phi)\rightarrow \left(t,\phi+\frac{2\pi}{\kappa_{\rm 1}}\right)\quad\hbox{or}\quad (t,\phi)\rightarrow \left(t,\phi+\frac{2\pi}{\kappa_{\rm 3}}\right),
\end{align}
should respectively be made. Since $\kappa_{\rm 1}\neq\kappa_{\rm 3}$ in general, we see that it is not possible to eliminate the conical singularities along both axes simultaneously. There is necessarily a conical singularity along at least one of the axes. Henceforth, we shall assume that the conical singularity along Rod 1 is eliminated. This results in a conical singularity with a deficit angle along Rod 3, pulling on the black hole.

\begin{figure}
 \begin{center}
  \includegraphics{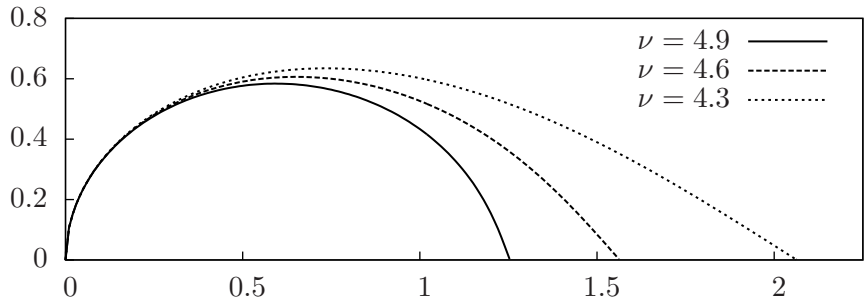}
  \caption{Examples of the horizon geometries of the deformed spherical black holes, as embeddings in three-dimensional Euclidean space. The units of the axes are defined only up to an overall scale. The full geometry can be visualised as the surface of revolution of these curves around the horizontal axis. The curves plotted are for $\mu=4$ and for $\nu$ values as indicated in the legend.}
  \label{embedding}
 \end{center}
\end{figure}

The geometry of the horizon represented by Rod 2 is described by the induced metric
\begin{align}
\label{induced}
 \dif s^2_{\rm H}=\frac{\ell^2}{x^2}\brac{\frac{\dif x^2}{G(x)}+G(x)\dif\phi^2},
\end{align}
where $x$ takes the range $-1\leq x\leq x_-$. To visualise this geometry, we can embed (\ref{induced}) as a surface of revolution in a three-dimensional Euclidean space (see, e.g., \cite{Gnecchi:2013mja}). Three examples of such a horizon geometry are plotted in Fig.~\ref{embedding} for a fixed $\mu$. As can be seen, the horizon takes the shape of a deformed sphere, with the amount of deformation increasing as $\nu$ decreases in value. The discontinuity that is present at the right pole of each sphere is the point at which the conical singularity touches the horizon.

The three rods form the left, upper and right edges of the trapezoidal domain respectively. Recall that the fourth edge of the trapezoid, given by $x=y$, corresponds to asymptotic infinity. This asymptotic region is locally AdS space, as can be seen from the Kretschmann invariant (\ref{Kret}). Indeed, if we approach this region along a line of constant $x$ and $\phi$, it can be checked that the metric (\ref{metric}) is asymptotically identical to that of AdS space in global coordinates.

We thus conclude that the space-times described by the trapezoidal domains of Region D2 contain a deformed spherical black hole, with a semi-infinite conical singularity attached to it. This black hole exists in an asymptotically AdS space; moreover, it is clearly static with respect to AdS infinity. Nevertheless, it feels the effect of the cosmological compression of AdS space, pushing it to the centre of the universe with a constant deceleration \cite{Krtous:2005ej}.\footnote{In the massless limit, the magnitude of the deceleration lies in the range $[0,\frac{1}{\ell})$, depending on the position of the black hole in AdS space.} It is the conical singularity which provides the necessary tension to counter this deceleration, keeping the black hole at a fixed position.

\subsection{Deformed hyperbolic black holes}
\label{sec5.2}

We now turn to the triangular domains of Regions A, B, C and D4. The left edge of the triangle represents a symmetry axis of the space-time, while the upper edge represents a horizon of the space-time. The remaining edge of the triangle represents asymptotic infinity. It follows that both the axis and horizon extend to asymptotic infinity in this case. In particular, the horizon does not have a compact topology; we shall see below that it is in fact asymptotically hyperbolic.

The rod structure of the space-time consists of the following two rods:
\begin{itemize}
 \item Rod 1: a semi-infinite space-like rod located at ($x=-1,-1<y\leq 0$), with direction $\ell_1=\frac{1}{\kappa_{\mathrm{1}}}\frac{\partial}{\partial\phi}$, where
  \begin{align}
   \kappa_{\mathrm{1}}=\half\brac{\mu-\nu+1}.
  \end{align}
  \item Rod 2: a semi-infinite time-like rod located at ($-1\leq x<0,y=0$), with direction $\ell_2=\frac{1}{\kappa_2}\frac{\partial}{\partial t}$, where
    \begin{align}
     \kappa_2=\half\brac{\nu+1}.
    \end{align}
\end{itemize}
Rod 1 is the asymptotic axis, while Rod 2 is the black-hole horizon that extends to asymptotic infinity. To avoid a conical singularity along Rod 1, one needs to impose the following periodicity on $\phi$:
\begin{align}
(t,\phi)\rightarrow \left(t,\phi+\frac{2\pi}{\kappa_1}\right).
\end{align}
The space-time is then completely regular along this axis.

The induced metric on the horizon represented by Rod 2 is again given by (\ref{induced}), except that $x$ now takes the entire range $-1\leq x<0$. Since $G(x)\rightarrow 1$ as $x\rightarrow 0$, the horizon geometry turns into (a quotient of) hyperbolic space in Poincar\'e coordinates when the horizon approaches infinity. This is confirmed by calculating the curvature invariants of this two-dimensional metric. The scalar curvature $R_{\rm H}$ and Kretschmann invariant $K_{\rm H}$ are
\begin{align}
  R_{\rm H}=-\frac{2}{\ell^2}\brac{1+\mu x^3},\qquad K_{\rm H}=R_{\rm H}^2\,. \label{curvatures}
\end{align}
Note that $R_{\rm H}$ is a negative constant when infinity is approached as $x\rightarrow0$. So we can think of this class of solutions as generically describing asymptotically hyperbolic black holes.
 
\begin{figure}
 \begin{center}
  \includegraphics[scale=1]{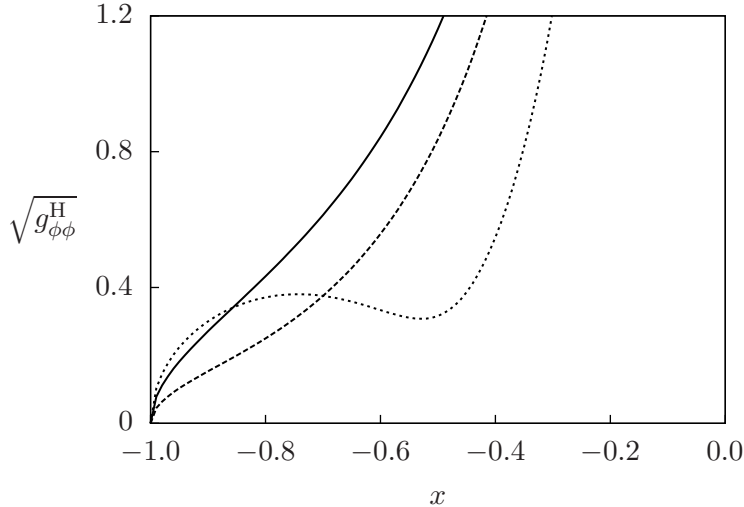}
  \caption{Examples of the horizon geometries of the deformed hyperbolic black holes, as plots of $\sqrt{g^{\rm H}_{\phi\phi}}$ versus $x$. The full geometry can be visualised as the surface of revolution of these curves around the $x$-axis. The solid, dashed and dotted curves have $(\mu,\nu)=(0.5,0.9)$, $(1,1.8)$ and $(4,3.9)$ respectively. The last case corresponds to a black globule. Since it has a positive scalar curvature, the globular part can actually be embedded in three-dimensional Euclidean space (c.f.~Fig.~\ref{embedding2}).}
  \label{horiz_geom}
 \end{center}
\end{figure}

For non-zero values of $x$, the geometry of the horizon will qualitatively depend on the value of $\mu$. From \Eqref{curvatures}, we can identify the three distinct cases:
\begin{itemize}
 \item $\mu<1$: The scalar curvature is always negative. From the $\mu$--$\nu$ plot of Fig.~\ref{fig2}, it can be seen that this case encompasses Regions A, B, C and part of D4.
 \item $\mu=1$: The scalar curvature is always negative except for a single point at $x=-1$, which has zero scalar curvature. This case lies entirely in Region D4.
 \item $\mu>1$: The scalar curvature is negative in a region sufficiently close to $x=0$, is positive in a region sufficiently close to $x=-1$, and is zero on a circle where these two regions meet. Again, this case lies entirely in Region D4.
\end{itemize}
Since the scalar curvature is negative in at least part of these geometries, we cannot fully embed them in a three-dimensional Euclidean space as in the previous subsection. To visualise these geometries, we can instead plot the proper length $\sqrt{g^{\rm H}_{\phi\phi}}$ as a function of $x$ \cite{Emparan:2009dj}. In Fig.~\ref{horiz_geom}, we have plotted representative examples of each of the three cases. In all the cases, the negativity of the scalar curvature as $x\rightarrow 0$ is manifest in the divergent nature of $\sqrt{g^{\rm H}_{\phi\phi}}$. 

What is more interesting however, is the manifestation of the positivity of the scalar curvature at the other end of the range, in the case when $\mu>1$. In this case, the horizon geometry becomes approximately spherical in a region centred around $x=-1$, resulting in a spherical protrusion out of the hyperbolic horizon. This spherical region is connected to the asymptotically hyperbolic region by a ``throat'', which can be as narrow as one desires by tuning the parameters $(\mu,\nu)$. Due to its resemblance to a drop of liquid, we shall call such a configuration a ``black globule''.\footnote{The alternative term ``black droplet'' might also be appropriate here, but we have noted that it has already been used in \cite{Hubeny:2009ru,Hubeny:2009kz} to describe a different type of black hole in AdS space.}

The two rods form the left and upper edges of the triangular domain respectively. The third edge of the triangle, given by $x=y$, corresponds to asymptotic AdS infinity. This region can be approached along a line of constant $x\neq0$ and $\phi$, similar to the case of the deformed spherical black holes. As in that case, the metric is asymptotically identical to that of AdS space in global coordinates.

We thus conclude that the space-times described by the triangular domains contain a deformed, asymptotically hyperbolic black hole. This black hole exists in an asymptotically AdS space, and is static with respect to AdS infinity. It is the hyperbolic analogue of the deformed spherical black holes discussed in the previous subsection. However, unlike the spherical case, there are no conical singularities present in this case. This deformed hyperbolic black hole is nonetheless able to remain static in the face of the cosmological compression of AdS space, since it is itself connected to AdS infinity. It is the horizon which provides the necessary tension to keep itself at a fixed position.

\section{Special cases}
\label{sec6}

\subsection{Non-deformed topological black holes}
\label{top_bh}

The usual topological black holes can be recovered from (\ref{metric}) by taking the limit
\begin{align}
\label{non-accel}
\nu\rightarrow 1+\mu\,,
\end{align}
in a suitable fashion. We first redefine the coordinates $(x,y)$ and parameters $(\mu,\nu)$ in terms of new coordinates $(r,\chi)$ and parameters $(\alpha,\epsilon)$ as
\begin{align}
\label{non-accel1}
x=-1+\epsilon \chi\,,\qquad y=-1+\frac{\ell}{\alpha r}\,,\qquad \mu=1+k\alpha^2,\qquad \nu=1+\mu-\epsilon\alpha^2,
\end{align}
where $k$ is a constant that takes values 0, $\pm1$. After rescaling $t$ and $\phi$ appropriately, and taking the limit $\epsilon\rightarrow 0$, we obtain the metric
\begin{subequations}\label{top_metric}
\begin{align}
\label{top_metric1}
\dif s^2&=-f(r)\dif t^2+f(r)^{-1}\dif r^2+r^2\left(\frac{\dif \chi^2}{\chi(1-k\chi)}+4\chi(1-k\chi)\dif \phi^2\right),
\end{align}
where
\begin{align}
f(r)&=k-\frac{2m}{r}+\frac{r^2}{\ell^2}\,,\qquad m\equiv\frac{\ell(1+k\alpha^2)}{2\alpha^3}\,.
\end{align}
\end{subequations}
The part of the metric enclosed by the large brackets in (\ref{top_metric1}) describes a two-dimensional space of constant positive, zero or negative curvature, for $k=+1,0,-1$ respectively. Thus, the metric (\ref{top_metric}) can be recognised to be that of the spherical Schwarzschild-AdS black hole for $k=+1$, the planar black hole for $k=0$, and the hyperbolic black hole for $k=-1$.

In the $\mu$--$\nu$ plot of Fig.~\ref{fig2}, this limit corresponds to approaching the semi-infinite line $\nu=1+\mu$, which forms the upper boundary of the shaded area. From the expression of $\mu$ in (\ref{non-accel1}), we see that the part of the line with $\mu>1$ corresponds to the class of spherical black holes. On the other hand, the point $\mu=1$ corresponds to the class of planar black holes, while the part with $-2<\mu<1$ corresponds to the class of hyperbolic black holes. Within the latter class, three distinct subclasses can be identified depending on the sign of the mass parameter $m$: The case of positive, zero and negative $m$ corresponds to the part of the line with $0<\mu<1$, $\mu=0$ and $-2<\mu<0$ respectively. In particular, the massless case $\mu=0$ is just AdS space in Rindler coordinates (see, e.g., \cite{Vanzo:1997gw,Emparan:1999gf}).

\begin{figure}
 \begin{center}
  \includegraphics[scale=0.6]{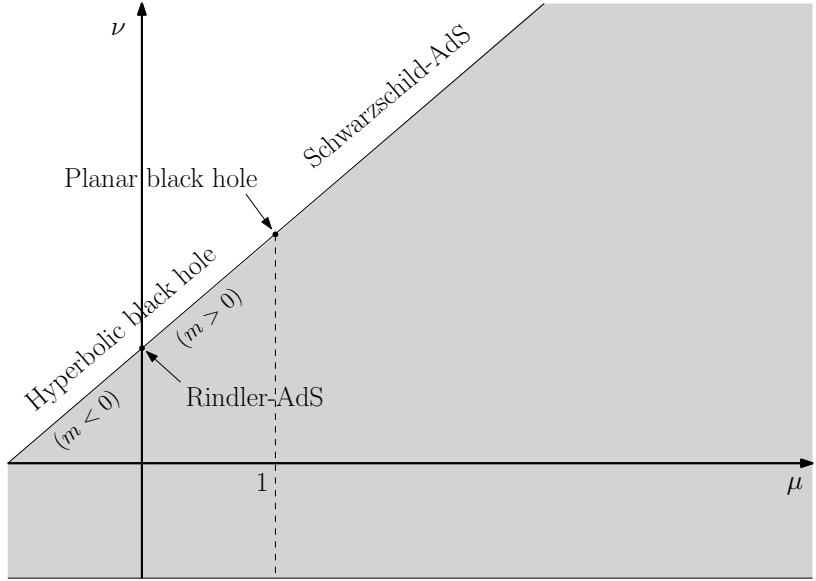}
  \medskip
  \caption{The interpretation of the various parts of the $\nu=1+\mu$ boundary of the $\mu$--$\nu$ plot of Fig.~\ref{fig2}, when the limit $\nu\rightarrow1+\mu$ is taken as described in the text.}
  \label{limits}
 \end{center}
\end{figure}

These identifications are summarised in Fig.~\ref{limits}. Note that they are consistent with our earlier interpretation of the regions immediately below the $\nu=1+\mu$ line. The region below the $\mu>1$ part of the line is Region D2, which we recall describes the class of ``slowly accelerating'' spherical black holes. The limit (\ref{non-accel}) then corresponds to turning off the acceleration, and hence deformation of the black holes, so that they become perfectly spherical. On the other hand, the regions below the $-2<\mu<0$ and $0<\mu<1$ parts of the line are Regions A and C respectively, which describe deformed hyperbolic black holes. The limit (\ref{non-accel}) again corresponds to turning off the deformation of the black holes, so that they become perfectly hyperbolic.

It is known that the hyperbolic black holes with positive and negative mass parameter $m$ have different causal structures \cite{Brill:1997mf}. The Penrose diagram for a hyperbolic black hole with positive $m$ is similar to that of a Schwarzschild-AdS black hole: there is a space-like singularity within the event horizon. For negative $m$ however, the Penrose diagram is similar to that of a Reissner--Nordstr\"om-AdS black hole: there is an inner horizon within the event horizon, and a time-like singularity within the former. These causal structures are actually consistent with those of the black holes described by Regions C and A respectively. For the black holes described by Region C, it can be seen from Fig.~\ref{figC} that there is always a space-like singularity within the event horizon. For the black holes described by Region A, it can be seen from Fig.~\ref{figA} that there is an inner horizon and a time-like singularity within the event horizon.

\subsection{Generalised Rindler-AdS space}

From the expression (\ref{Kret}) of the Kretschmann invariant of the general space-time, we see that the latter must be locally isometric to AdS space if $\mu=0$. We have already seen a special case of this in the previous subsection: a suitable limit in which $\nu\rightarrow1$ is AdS space in Rindler coordinates. In this subsection, we will investigate the case where $\nu$ can take values in the range $(-1,1)$.

\begin{figure}
\begin{center}
 \begin{subfigure}[b]{0.4\textwidth}
  \centering
  \includegraphics[scale=0.45]{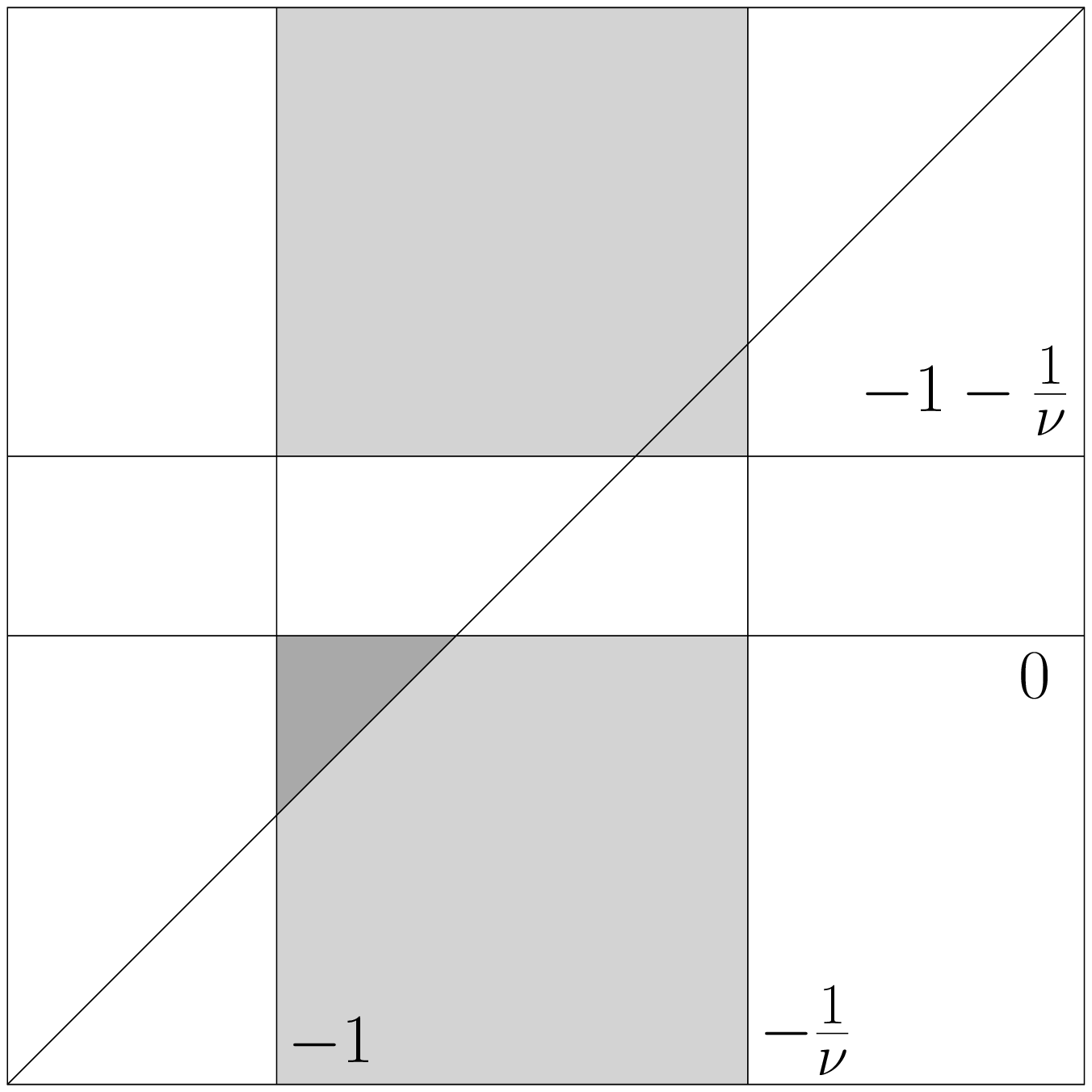}
  \caption{$-1<\nu<0$}
  \label{fig4a}
 \end{subfigure}\hspace{12pt}
 \begin{subfigure}[b]{0.4\textwidth}
  \centering
  \includegraphics[scale=0.45]{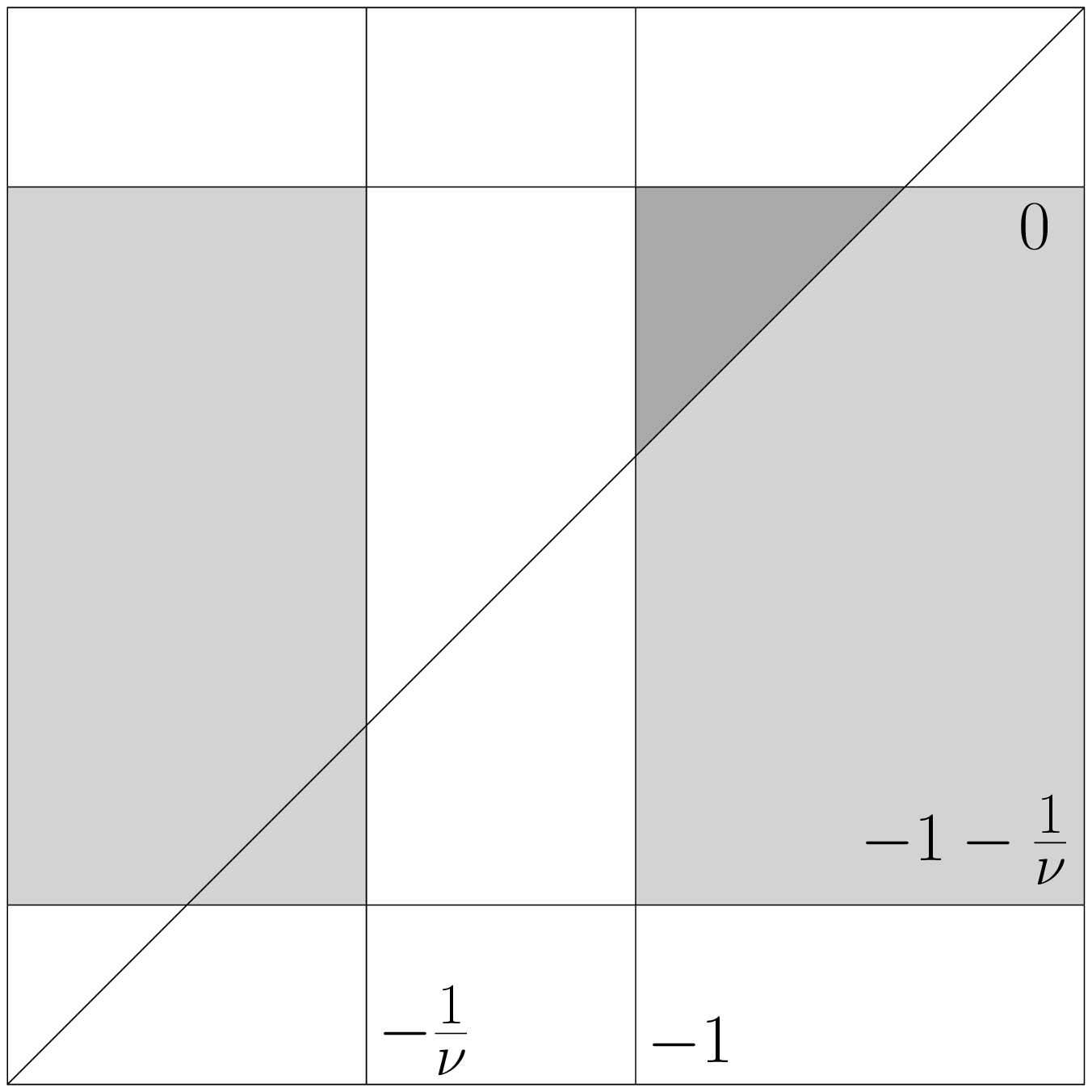}
  \caption{$0<\nu<1$}
  \label{fig4b}
 \end{subfigure}
\end{center}
\caption{The domain structures for the case $\mu=0$ with (a) $-1<\nu<0$; and (b) $0<\nu<1$.}
\label{fig4}
\end{figure}

If we set $\mu=0$, the metric (\ref{metric}) reduces to
\begin{align}
\label{gen_Rindler}
 \dif s^2&\eq\frac{\ell^2}{(x-y)^2}\brac{F(y)\dif t^2-\frac{\dif y^2}{F(y)}+\frac{\dif x^2}{G(x)}+G(x)\dif\phi^2},\nonumber\\
    F(y)&\eq y(1+\nu+\nu y)\,,\qquad G(x)\eq(1+x)(1+\nu x)\,.
\end{align}
Note that the structure functions are now quadratic polynomials, whose roots can be trivially read off. As usual, we consider the coordinate range $-1<x<y<0$. Depending on the sign of $\nu$, there are now two possible domain structures as depicted in Fig.~\ref{fig4}. In either case, the coordinate range of interest forms a triangular domain. Thus the space-times described by (\ref{gen_Rindler}) contain an axis and a horizon, both extending to asymptotic infinity. 

As mentioned, the space-times described by (\ref{gen_Rindler}) must be locally isometric to AdS space. This metric can be cast in a more familiar form of AdS space by introducing the transformation
\begin{align}
 x\rightarrow\frac{\ell x}{\sqrt{\left|\nu\right|}}-\frac{\nu+1}{2\nu}\,,\qquad y\rightarrow\frac{\ell y}{\sqrt{\left|\nu\right|}}-\frac{\nu+1}{2\nu}\,,\qquad t\rightarrow\frac{\sqrt{\left|\nu\right|}t}{\ell}\,,\qquad\phi\rightarrow\frac{\sqrt{\left|\nu\right|}\phi}{\ell}\,.
\end{align}
It then becomes
\begin{align}
 \dif s^2&\eq\frac{1}{(x-y)^2}\brac{{F(y)}\dif t^2-\frac{\dif y^2}{{F(y)}}+\frac{\dif x^2}{{G(x)}}+{G(x)}\dif\phi^2},\nonumber\\
 {F(y)}&\eq\frac{\nu}{\left|\nu\right|}y^2-\frac{(\nu+1)^2}{4\ell^2\nu}\,,\qquad{G(x)}=\frac{\nu}{\left|\nu\right|}x^2-\frac{(\nu-1)^2}{4\ell^2\nu}\,. \label{accel_tpbh}
\end{align}
When $-1<\nu<0$, this metric is known to arise when the massless limit of the usual AdS C-metric is taken. Indeed, if we define the parameter $A$ by
\begin{align}
\label{A}
 A^2=-\frac{(\nu-1)^2}{4\ell^2\nu}\,,
\end{align}
then (\ref{accel_tpbh}) is equivalent to the space-time considered in Sec.~IV.A of \cite{Dias:2002mi} after a rescaling of the coordinates.

This space-time with $-1<\nu<0$ was analysed in detail in \cite{Dias:2002mi}, where it was shown to describe AdS space in a certain set of accelerated coordinates. In particular, it was shown that $A$ is simply the proper acceleration of an observer at the point $y=+\infty$ of the space-time. This point is special in that if a (spherical) black hole is reintroduced into the space-time, the source of the black hole---its singularity---is located there. The parameter $A$ can thus be identified as the acceleration of the black hole, at least in the weak-field limit.

It should be emphasized that the space-times described by (\ref{accel_tpbh}) are generally different from Rindler-AdS space, which also describes AdS space in a set of accelerated coordinates. Recall that the latter is foliated by hyperbolic 2-spaces of maximal symmetry \cite{Vanzo:1997gw}. However, the former space-times are foliated by surfaces of reduced symmetry, in this case {\it deformed\/} hyperbolic spaces. For this reason, we shall refer to them as generalised Rindler-AdS spaces.

If a hyperbolic black hole is reintroduced into such a space-time, it is therefore not surprising that its horizon takes the form of a deformed hyperbolic space. These black holes are deformed analogues of the usual hyperbolic black holes that are obtained when a mass is introduced to Rindler-AdS space (c.f.\ Sec.~\ref{top_bh}). 

There is a certain sense in which the parameter $A$, given by (\ref{A}), can be identified as the acceleration of these deformed hyperbolic black holes. If such a black hole is reintroduced into the generalised Rindler-AdS spaces by restoring the parameter $\mu$ in (\ref{gen_Rindler}), a curvature singularity will appear inside the event horizon, at $y=+\infty$. Identifying this singularity as the source of the black hole, we can then conclude that it has a constant proper acceleration $A$, at least in the weak-field limit. Recall that this calculation of the acceleration was done \cite{Dias:2002mi} assuming the observer at $y=+\infty$ is time-like, which would be the case if $-1<\nu<0$. On the other hand, if $0<\nu<1$, the observer will have a space-like 4-velocity and hence a time-like 4-acceleration. Nevertheless, it can be checked that the magnitude of this 4-acceleration will continue to be given by (\ref{A}). In this sense, $A$ can still be interpreted as the acceleration of the black hole.

\subsection{Black funnels}
\label{funnels}

In Sec.~\ref{top_bh}, we took the limit 
\begin{align}
\label{bf}
\nu\rightarrow 1+\mu\,,
\end{align}
while simultaneously sending $x\rightarrow-1$ in a suitable fashion. In this subsection, we shall consider taking the limit (\ref{bf}) directly. In this case, $G(x)$ will have a double root at $x=-1$. This arises from the coalescence of the two roots $x_-$ and $x_0$ of (\ref{G_roots}) in the case $\mu>1$, and from the coalescence of the two roots $x_+$ and $x_0$ in the case $\mu<1$.

Now in Sec.~\ref{sec3}, we found all possible orderings of the roots $x_0$, $x_\pm$, $y_0$ and $y_\pm$ that give rise to Lorentzian space-times. To preserve the orderings, two roots of $G(x)$ are allowed to coalesce only if they are real and adjacent to each other. Note that the coalescence of $x_-$ and $x_0$ is allowed only if the roots take the ordering (\ref{order_D}), and if $y_\pm$ are imaginary. This corresponds to Region D2.\footnote{\label{foot}Regions D4 and C3 are excluded since $x_\pm$ are complex in these cases.} On the other hand, the coalescence of $x_+$ and $x_0$ is allowed only if the roots take the ordering (\ref{order_A}) or (\ref{order_C}). The former ordering corresponds to Region A, while the latter corresponds to Region C1 or C2.$^{\ref{foot}}$ As a consistency check, we note that all these regions border the $\nu=1+\mu$ line in Fig.~\ref{fig2}. 

The coalescence of the roots can be visualised in terms of the domain structures of the various regions. We see from the domain structure of Region D2 in Fig.~\ref{figD2} that the limit $x_-\rightarrow-1$ will actually shrink the darker-shaded trapezoidal region down to zero width. This is a trivial limit unless we take $x\rightarrow-1$ simultaneously, as was done in Sec.~\ref{top_bh}. Thus we will not consider this case any further. 

If we take the limit $x_+\rightarrow-1$ in the domain structure of Region A in Fig.~\ref{figA}, the darker-shaded triangular region will appear to merge with the lighter-shaded region to its left. However, these two regions will remain physically distinct, because the double root at $x=-1$ will make this line infinitely far away from other points of the space-time. Thus the $x=-1$ line becomes part of infinity of the space-time just like the $x=y$ line, and the domain preserves its triangular shape. Similar remarks apply to the domain structures of Regions C1 and C2 in Fig.~\ref{figC}.

In each of these cases, the triangular domain describes a space-time with a horizon stretching between asymptotic infinity at $x=y$, and the new spatial infinity at $x=-1$. Such space-times are just the original black funnel solutions of \cite{Hubeny:2009ru}. We remark that the subsequent definition of black funnels in \cite{Hubeny:2009kz} does not require the existence of the new spatial infinity, so any triangular domain would describe a black funnel by this definition.

\subsection{A class of black holes with unusual horizons}

A double root for $G(x)$ will also occur if the limit
\begin{align}
\label{sebh}
\nu\rightarrow \pm2\sqrt{\mu}\,,
\end{align}
is taken. This arises from the coalescence of the two roots $x_+$ and $x_-$ of (\ref{G_roots}). In this subsection, we would like to see if the resulting double root can give rise to a new spatial infinity of the space-time, as in the case considered in the previous subsection. To do so, we turn directly to the domain structures of the various regions.

It can be seen that the coalescence of $x_+$ and $x_-$ is allowed only in Regions B1, C1, C2, D1 and D2. In the former three cases however, the double root at $x=x_\pm$ remains separated from the domain of interest, and will not have a direct effect on it. It is only in Regions D1 and D2 that the double root will border the domain of interest, giving rise to a new spatial infinity of the space-time. 

Since Regions D1 and D2 lie in the part of the parameter range with positive $\nu$, the coalescence limit corresponds to taking the upper sign in (\ref{sebh}). In the context of Fig.~\ref{fig2}, it corresponds to approaching the $\nu=2\sqrt{\mu}$ curve from either region. We also recall that the shapes of the respective domains will be preserved in this limit. In particular, Region D1 will continue to have a box-shaped domain. This case was briefly analysed in \cite{Chen:2015vma}, where it was identified as the original black droplet solution of \cite{Hubeny:2009ru}. In this subsection, we will focus on the trapezoidal domain of Region D2.

\begin{figure}
 \begin{center}
  \includegraphics[scale=1]{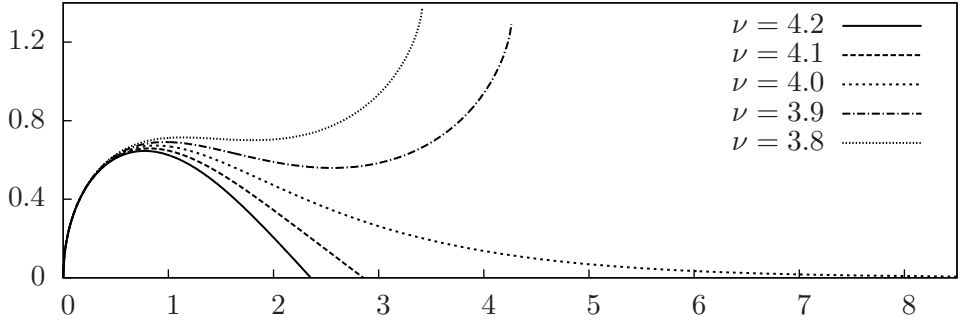}
  \caption{Examples of horizon geometries for $\mu=4$ and for values of $\nu$ at and around the critical value $2\sqrt{\mu}$, as embeddings in three-dimensional Euclidean space. As usual, the full geometry can be visualised as the surface of revolution of these curves around the horizontal axis.}
  \label{embedding2}
 \end{center}
\end{figure}

Recall that the generic trapezoidal domain of Region D2 was analysed in detail in Sec.~\ref{sec5.1}, where it was shown to describe a class of deformed spherical black holes. The limit (\ref{sebh}) (with upper sign) can then be taken within this class. It is clear from the sequence of plots in Fig.~\ref{embedding} that for fixed $\mu$, the deformation of the black-hole horizon increases as $\nu$ is decreased. This sequence is continued in Fig.~\ref{embedding2}. The plots show that the horizon becomes more elongated, and in fact becomes infinite in extent when $\nu$ reaches the critical value $2\sqrt{\mu}$. This is consistent with the above-mentioned fact that $x=x_-$ is now infinitely far away.

In Fig.~\ref{embedding2}, we have also plotted examples of horizon geometries in the case when $\nu$ is decreased {\it below\/} the critical value $2\sqrt{\mu}$. In this case, we enter the part of Region D4 whose space-times contain the black globules first discussed in Sec.~\ref{sec5.2}. Thus the black holes with critical value $\nu=2\sqrt{\mu}$ can alternatively be obtained as a limiting case of black globules in which the throat becomes infinitely long and narrow. 

Perhaps the most remarkable property of the black holes in the critical case is that although their horizons are infinite in extent, they have {\it finite\/} area. This can be verified from the induced metric (\ref{induced}) on the horizon, and is due to the fact that $g^{\rm H}_{\phi\phi}\rightarrow0$ sufficiently rapidly when infinity is approached as $x\rightarrow x_-$. In this sense $x=x_-$ can be regarded as an axis, although it is actually excised from the space-time. The result is a ``puncture'' in the horizon at infinity. The topology of the horizon is thus that of a sphere with one puncture. Moreover, the space-time is completely regular outside the horizon.

Black holes with this type of horizon are reminiscent of a new class of black holes recently discovered by Klemm et al.~\cite{Gnecchi:2013mja,Klemm:2014rda}, and whose properties have been further studied in \cite{Hennigar:2014cfa,Hennigar:2015cja}. Like the above black holes, they have finite area even though they are infinite in extent. The main difference is that Klemm's black holes are stationary, and have horizons that are topologically spheres with {\it two\/} punctures. It would be interesting to clarify the relationship between the static black holes considered here and Klemm's black holes, and to study their properties in more detail.

Finally, we mention that this black hole can be also recovered from the original black droplet solution of \cite{Hubeny:2009ru}. Recall that this solution has two disconnected horizons: one extending to asymptotic infinity $x=y$, and the other extending to the new spatial infinity $x=x_-$. If the former black hole is pushed beyond extremality so that it actually disappears (instead of leaving a naked singularity), then the remaining black hole is precisely the one considered here.

\subsection{Extremal deformed hyperbolic black holes}

It is also possible to consider the situation in which $F(y)$ has a double root at $y=0$. This occurs in the limit
\begin{align}
\label{extremal}
\nu\rightarrow-1\,,
\end{align}
which results in the coalescence of the two roots $y_+$ and  $y_0$ of (\ref{F_roots}) in the case $\mu>1$, and in the coalescence of the two roots $y_-$ and  $y_0$ in the case $\mu<1$. In the context of Fig.~\ref{fig2}, this limit corresponds to approaching the lower boundary of the shaded area, from Region D3 if $\mu>1$ and from Region A, B1 or B2 if $\mu<1$.

As before, the coalescence of the roots can be visualised in terms of the domain structures of these regions. We see from the domain structure of Region D3 in Fig.~\ref{figD3} that the limit $y_+\rightarrow0$ will actually shrink the darker-shaded trapezoidal region down to zero width. Thus we will not consider this case any further. 

On the other hand, if we take the limit $y_-\rightarrow0$ in the domain structure for Region A, B1 or B2, we see that the darker-shaded triangular region will join up with the lighter-shaded region above it. This can be interpreted as the black-hole horizon merging with the inner horizon and becoming degenerate, and is consistent with the fact that the temperature of the black-hole horizon $T=\frac{\kappa_2}{2\pi}$ vanishes in this limit. Thus, the solutions obtained in the limit (\ref{extremal}) for $\mu<1$ can be interpreted as a class of extremal deformed hyperbolic black holes.

The reader may have noticed a certain similarity between the domain structures of these extremal black holes, and those of the black funnels of Sec.~\ref{funnels}. Indeed, they are related by the transformation (\ref{discrete}), which recall effectively flips the domain structure about the diagonal line joining the upper-left and lower-right corners. The solution describing these extremal black holes is therefore related to the black-funnel solution by the double-Wick rotation $t\rightarrow i\phi$ and $\phi\rightarrow it$.

\section{Relation to traditional forms}
\label{sec7}

The AdS C-metric describing accelerating topological black holes is traditionally written in the general form (see, e.g., \cite{Mann:1996gj}):
\begin{align}
 \dif s^2&=\frac{1}{\tilde A^2(\tilde x-\tilde y)^2}\brac{\tilde F(\tilde y)\dif \tilde t^2-\frac{\dif \tilde y^2}{\tilde F(\tilde y)}+\frac{\dif \tilde x^2}{\tilde G(\tilde x)}+\tilde G(\tilde x)\dif\tilde \phi^2},\nonumber\\
      \tilde G(\tilde x)&=\tilde \gamma-\tilde b\tilde x^2-2\tilde m\tilde A\tilde x^3,\qquad
\tilde F(\tilde y)=\tilde \lambda-\tilde b\tilde y^2-2\tilde m\tilde A\tilde y^3,\label{Mann_form}
\end{align}
where $\tilde\lambda\equiv\tilde\gamma-\frac{1}{\ell^2\tilde A^2}$. Here, $\tilde\gamma$ and $\tilde b$ are kinematical parameters whose values can be fixed arbitrarily up to a sign. They are usually taken to have values $\pm1$ or $0$. The standard AdS C-metric describing accelerating spherical black holes is obtained with the choice $\tilde\gamma=1$ and $\tilde b=1$. On the other hand, the choice $\tilde\gamma=1$ and $\tilde b=0$ describes an accelerating planar black hole, while the choice $\tilde\gamma=-1$ and $\tilde b=-1$ describes an accelerating hyperbolic black hole. How these black holes can be recovered in the non-accelerating limit is described in \cite{Mann:1996gj}.

We can get from \Eqref{metric} to \Eqref{Mann_form} by considering the transformation
\begin{align}
 x=Bc_0\tilde x+c_1\,,\qquad y=Bc_0\tilde y+c_1\,,\qquad t=\frac{c_0}{B}\,\tilde t\,,\qquad\phi=\frac{c_0}{B}\,\tilde\phi\,,
\end{align}
where $B$, $c_0$ and $c_1$ are (real) constants to be determined. To preserve the form of the metric, we require that
\begin{align}
 \frac{\ell^2}{B^2}=\frac{1}{\tilde A^2}\,,\qquad \frac{G(x)}{B^2}=\tilde G(\tilde x)\,,\qquad \frac{F(y)}{B^2}=\tilde F(\tilde y)\,. \label{require}
\end{align}
Equating the coefficients of the structure functions in \Eqref{require}, we obtain
\begin{subequations}
\begin{align}
 -2\tilde m\tilde A&=Bc_0^3\mu\,,\label{x3}\\
 -\tilde b&=c_0^2(\mu+\nu+3\mu c_1)\,,\label{x2}\\
 0&=3\mu c_1^2+2(\mu+\nu)c_1+\nu+1\,,\label{x1}\\
 \tilde\gamma&=\frac{1}{B^2}(1+c_1)(1+\nu c_1+\mu c_1^2)\,,\label{x0}\\
 \tilde\lambda&=\frac{c_1}{B^2}\brac{1+\nu+(\mu+\nu)c_1+\mu c_1^2}.\label{y0}
\end{align}
\end{subequations}
Note that (\ref{y0}) is not an independent equation, but can be obtained from (\ref{x0}) and the first equation of (\ref{require}).

\begin{figure}
\begin{center}
  \includegraphics[scale=0.6]{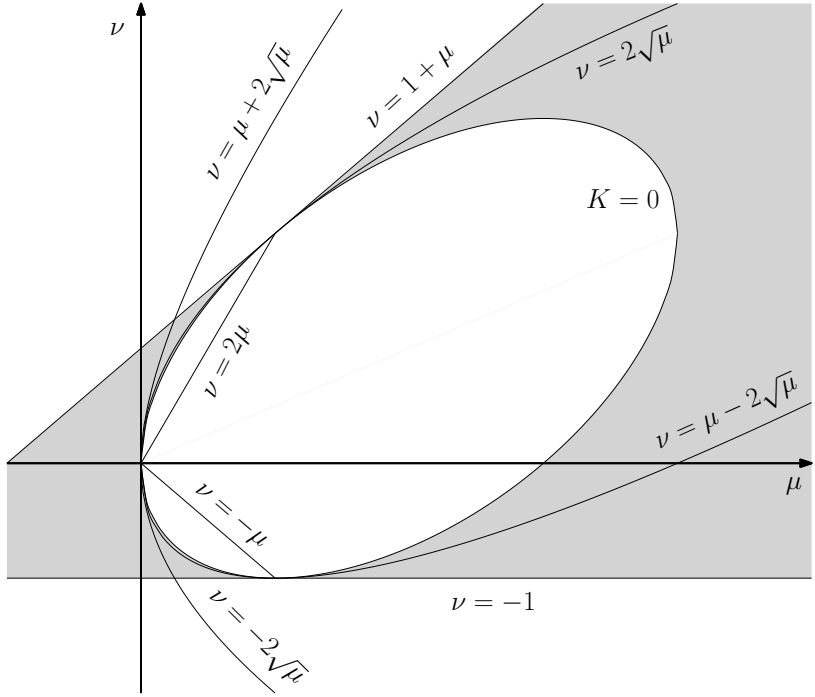}
\end{center}
\caption{The closed curve $K=0$ in the $\mu$--$\nu$ plot of Fig.~\ref{fig2}. The shaded regions outside this curve with $K>0$ are parameters which have a correspondence to those in the traditional form (\ref{Mann_form}). The labelling of the shaded regions follows that of Fig.~\ref{fig2}.}
\label{traditional}
\end{figure}

Solving the quadratic equation for $c_1$ in \Eqref{x1}, we have
\begin{align}
 c_1=\frac{-\mu-\nu\pm\sqrt{K}}{3\mu}\,, \label{c1_eqn}
\end{align}
where we have defined 
\begin{align}
K\equiv\mu^2+\nu^2-\mu\nu-3\mu\,. 
\end{align}
Substituting \Eqref{c1_eqn} into \Eqref{x2} then gives
\begin{align}
 \label{c0_eqn}
 -\tilde b=\pm c_0^2\sqrt{K}\,.
\end{align}
It follows that the $\tilde b=0$ case corresponds to $K=0$, whose solution is given by
\begin{align}
\label{K=0}
 \nu=\half\brac{\mu\pm\sqrt{12\mu-3\mu^2}},
\end{align}
and traces out a closed curve in a $\mu$--$\nu$ plot. As can be seen from Fig.~\ref{traditional}, this $K=0$ curve lies in Regions B3, C3 and D4. In particular, it passes through the point $(\mu,\nu)=(1,2)$, which recall corresponds to the planar black hole in a suitable limit. This is consistent with the known fact that the $\tilde b=0$ case contains this black hole in the non-accelerating limit. 

In the case when $\tilde b$ is non-zero, we need to choose the upper sign in \Eqref{c0_eqn} for $\tilde b=-1$ and the lower sign for $\tilde b=+1$, in order to have a real solution for $c_0$. At the same time, we require that $K>0$, which restricts us to the region of the $\mu$--$\nu$ plot {\it outside\/} the $K=0$ curve. In other words, the region inside the $K=0$ curve does not have a correspondence to parameters in the traditional form. This shows that the new form (\ref{metric}) of the AdS C-metric proposed in this paper is in fact more general than \Eqref{Mann_form}.\footnote{There is a direct way to see why this is so. In traditional forms of the C-metric, such as (\ref{Mann_form}), the linear coefficient in the structure functions is set to zero. This presumes that the structure functions have a stationary point, which is not always guaranteed for cubic polynomials. In fact, the condition $K<0$ is precisely that for the structure functions (\ref{metric2}) to have {\it no\/} stationary points.}

At this stage, we can read off the solution for $B$ from (\ref{x3}). Substituting the solutions for $B$, $c_0$ and $c_1$ into \Eqref{x0} and \Eqref{y0}, we obtain for $\tilde b=\mp1$,
\begin{subequations}\label{gamma-lambda}
\begin{align}
 4\tilde m^2\tilde A^2\tilde\gamma &=-\frac{1}{27K^{3/2}}\brac{2\mu-\nu\pm\sqrt{K}}^2\brac{-2\mu+\nu\pm2\sqrt{K}},\label{gamma}\\
4\tilde m^2\tilde A^2\tilde\lambda&=-\frac{1}{27K^{3/2}}\brac{-\mu-\nu\pm\sqrt{K}}^2\brac{\mu+\nu\pm2\sqrt{K}}.\label{lambda}
\end{align}
\end{subequations}
For $\tilde b=0$ however, we cannot solve (\ref{c0_eqn}) for $c_0$. Instead, we have a solution for $\nu$ given by (\ref{K=0}). Substituting this and the solutions for $B$ and $c_1$ into \Eqref{x0} and \Eqref{y0}, we obtain
\begin{subequations}\label{gamma-lambda1}
\begin{align}
 4\tilde m^2\tilde A^2\tilde\gamma &=\frac{c_0^6}{27}(2\mu-\nu)^3,\label{gamma1}\\
4\tilde m^2\tilde A^2\tilde\lambda&=-\frac{c_0^6}{27}(\mu+\nu)^3.\label{lambda1}
\end{align}
\end{subequations}
For fixed $\tilde\gamma$ and $\tilde b$, the equations (\ref{gamma-lambda}) or (\ref{gamma-lambda1}) can be used to establish a correspondence between the parameters in the new form (namely $\mu$ and $\nu$) and those in the traditional form (namely $\tilde m$ and $\tilde A$). 

For example, taking $\tilde b=+1$, it can be checked that the right-hand side of (\ref{gamma}) (with lower signs) is positive in almost all the shaded regions of the $\mu$--$\nu$ plot in Fig.~\ref{traditional}, except the thin sliver that is Region C3. Thus we can set $\tilde\gamma=+1$ in these regions and $\tilde\gamma=-1$ in Region C3. A similar argument for the right-hand side of (\ref{lambda}) shows that $\tilde\lambda$ is positive in Regions A, B, C1, D1 and D3, and is negative in the remaining regions. 

\begin{table}[!t]
\begin{center}
{\renewcommand{\arraystretch}{1.2}
\begin{tabular}{|r|r|l|}
  \hline
$\tilde\gamma$~ & $\tilde b$~ & ~Regions \\
\hline
1~ &  1~ & ~A, B, C1, C2, D\\
$-1$~ &  1~ & ~C3 \\
1~ &  $-1$~ & ~B2, B3, D3, D4\\
$-1$~ &  $-1$~ & ~A, B1, C, D1, D2~ \\
1~ &  0~ & ~B3, D4\\
$-1$~ & 0~ & ~C3 \\
\hline
\end{tabular}
}
\caption{Possible values for $\tilde\gamma$ and $\tilde b$, and the regions of the $\mu$--$\nu$ plot that they map to. For $\tilde b=\pm1$, the regions B3, C3 and D4 are understood to exclude the part $K\leq0$. For $\tilde b=0$ however, these regions are understood to lie on the curve $K=0$ itself.}
\label{table}
\end{center}
\end{table}

Table \ref{table} lists the possible values for $\tilde\gamma$ and $\tilde b$, and the regions of the $\mu$--$\nu$ plot that they map to.\footnote{For simplicity, we have not considered the possibility that $\tilde\gamma=0$. It can be checked that such cases correspond to the boundaries between the various regions listed in the table.} It can be seen that the standard choice $\tilde\gamma=1$ and $\tilde b=1$ actually maps to all the regions except C3 (and of course, the parts of B3 and D4 where $K\leq0$). In particular, Region D2 which describes the class of ``slowly accelerating'' spherical black holes is included.\footnote{We have noted above that $\tilde\lambda$ is negative in this region. This is equivalent to the condition $\tilde A^2<\frac{1}{\ell^2}$ that such black holes are known to satisfy.} Regions A, C1 and C2 describing accelerating hyperbolic black holes are also included, although they are more traditionally associated with the choice $\tilde\gamma=-1$ and $\tilde b=-1$.

The choice $\tilde\gamma=1$ and $\tilde b=-1$ has also been used to describe accelerating hyperbolic black holes (see, e.g., \cite{Emparan:1999fd,Hubeny:2009kz}). It can be seen from Table \ref{table} that this choice maps to Regions B2, B3, D3 and D4, which share the common property that $G(x)$ has only one real root. Since these regions are not connected to the $\nu=1+\mu$ line, the black-hole solutions they describe do not possess the zero-acceleration limit (\ref{non-accel}).

It is sometimes thought that the entire class of AdS C-metric solutions can be obtained from (\ref{Mann_form}) by setting $\tilde\gamma=1$, and considering the different cases $\tilde b=\pm1$ and 0. This is incorrect on two counts. Firstly, as we have already explained, an entire region $K<0$ of the $\mu$--$\nu$ parameter space is left out in this parameterisation. Secondly, as can be seen from Table \ref{table}, these values of $\tilde\gamma$ and $\tilde b$ do not map to all the possible regions with $K\geq0$. Indeed, Region C3 with $K\geq0$ is left out in this parameterisation. It is only included if the case $\tilde\gamma=-1$ is considered.

\section{Summary and discussion}

In this paper, we have presented a new form of the AdS C-metric (\ref{metric}), whose structure functions are assumed to have at least one real root each. It is more general than the form used in \cite{Chen:2015vma}, which presumes the existence of at least two real roots for each structure function. We have also shown that it is more general than the traditional forms of the AdS C-metric used in say \cite{Mann:1996gj,Emparan:1999fd,Hubeny:2009kz}.

We then found the complete range of parameters for which this metric describes a Lorentzian space-time. This parameter range, which is shown in Fig.~\ref{fig2}, can be divided into a number of subregions, each of which has a different domain structure. The shapes that the domains can take are boxes, trapezoids and triangles. The box domains were already studied in our previous paper \cite{Chen:2015vma}. In this paper, we focussed on the trapezoidal domain of Region D2, and the triangular domains arising from various different regions. 

The trapezoidal domain turns out to describe a deformed spherical black hole in AdS space. Although the black hole undergoes a constant acceleration, there is no acceleration horizon in the space-time. This space-time has been studied previously \cite{Podolsky:2002nk,Dias:2002mi,Krtous:2005ej}, and is now understood to describe a black hole that remains static with respect to AdS infinity. A conical singularity attached to the black hole provides the necessary tension to counterbalance the cosmological compression of AdS space.

The main focus of this paper however, was on the less well understood triangular domains. We have shown that these space-times contain an asymptotically hyperbolic black hole. They generalise the usual hyperbolic black hole \cite{Mann:1996gj,Vanzo:1997gw,Brill:1997mf,Mann:1997iz,Birmingham:1998nr}, in the sense that the horizon is now a deformed hyperbolic space. Indeed, we have found that the deformation can be quite non-trivial: in some cases leading to the formation of a spherical protrusion (Fig.~\ref{horiz_geom}), which we call a black globule. These black holes can be regarded as hyperbolic analogues of the deformed spherical black holes. One difference between them however, is that there are no conical singularities attached to the former. The deformed hyperbolic black holes can remain static in AdS space, since they are themselves connected to AdS infinity.

We have also found that a certain limit of the black globules can be taken, such that its throat becomes infinitely long and narrow. This limit (the middle curve in Fig.~\ref{embedding2}) results in a horizon which has finite area, even though it is infinite in extent. Its topology is that of a sphere with one puncture. This black hole resembles a class of stationary black holes recently found in \cite{Gnecchi:2013mja,Klemm:2014rda}, whose areas are also finite even though they are infinite in extent. However, their horizons are topologically spheres with two punctures instead of one.

Although we have not analysed the trapezoidal domain of Region D3 in this paper, it is possible to come up with a tentative physical interpretation of this space-time. Recall that Regions D2 and D1 describe spherical black holes, with the difference between them being that the black holes of D2 have an acceleration smaller than a certain critical value, while those of D1 have an acceleration larger than this critical value. We have seen that the former black holes are effectively static with respect to AdS infinity. On the other hand, the latter black holes are truly accelerating, resulting in the appearance of an acceleration horizon in the space-time. Turning to Region D4, we have seen that it describes hyperbolic analogues of the spherical black holes of Region D2. Because there is an additional horizon in Region D3 compared to D4, we can then conclude that Region D3 describes hyperbolic black holes undergoing a sufficiently large acceleration, such that an acceleration horizon appears in the space-time.

In this paper, we have mostly confined ourselves to the static region of the space-time outside the hyperbolic black-hole horizon. To describe the region of the space-time inside the horizon, we need to use non-static coordinates that analytically extend past it. In the context of the domain structures of Figs.~\ref{figA}--\ref{figD}, this corresponds to entering the region above the dark-shaded domain. It can be seen that there is necessarily a curvature singularity in this region at $y=+\infty$, although the observer may have to cross one or more inner horizons before reaching it. We mention that the maximally extended space-time can be quite non-trivial in certain cases, and may lead to alternative interpretations for the hyperbolic black-hole horizon. For example, the space-time past the horizon in Fig.~\ref{figB1} actually contains the accelerating black-hole space-time described by Fig.~\ref{figD1}. This maximally extended space-time was constructed in \cite{Krtous:2005ej}, and the horizon in question was interpreted as a cosmological horizon separating pairs of accelerating black holes. One interesting avenue for future research would be to study the maximal analytic extensions of the various space-times described by (\ref{metric}) in more detail. 

We remark that it is straightforward to add an electric charge $e$ and a magnetic charge $g$ to the solution (\ref{metric}). The resulting metric is given by
\begin{align}
\dif s^2&=\frac{\ell^2}{(x-y)^2}\left(F(y)\dif t^2-\frac{\dif y^2}{{F(y)}}+\frac{\dif x^2}{{G(x)}}+{G(x)}\dif \phi^2\right),\cr
F(y)&=y\sbrac{1+\nu+(\mu+\nu)y+(\mu-q^2)y^2-q^2y^3},\cr
G(x)&=(1+x)\brac{1+\nu x+\mu x^2-q^2x^3},
\end{align}
where $q^2\equiv e^2+g^2$. The corresponding gauge potential is
\begin{align}
{\cal A}=\ell(ey\dif t-g(1+x)\dif \phi)\,.
\end{align}
Note that $F(y)$ and $G(x)$ are now quartic polynomials; nevertheless, the analysis of the uncharged case can readily be extended to this case. In analysing the possible domain structures, we again find trapezoidal and triangular domains. They describe charged generalisations of the deformed spherical and hyperbolic black holes respectively. A subclass of the charged deformed hyperbolic black holes has previously been studied in \cite{Caldarelli:2011wa}.

Another possible generalisation of the solution (\ref{metric}) is to include rotation and NUT charge. Such a solution can be extracted from the general Pleba\'nski--Demia\'nski solution \cite{Plebanski:1976gy}, through the systematic analysis of the possible domain structures arising from the latter. In particular, there would no doubt exist a plethora of different triangular domains, all of which would describe rotating and/or NUT-charged hyperbolic black holes. We only remark here that NUT-charge has a different physical significance for hyperbolic black holes, as compared to spherical ones. Recall that for a spherical black hole, the presence of NUT charge will necessarily lead to closed time-like curves (CTCs) in the space-time. This occurs because the symmetry axis has been separated into two parts by the black hole. Roughly speaking, the NUT charge will present an obstacle to finding a global identification of the angular coordinate, that would make both axes regular simultaneously. The best one can do is to make one axis regular, at the expense of having CTCs around the other. This problem does not exist for hyperbolic black holes however, since there is only one axis in the space-time. It remains an outstanding problem to understand the true physical significance of NUT charge for such black holes.

It might also be worthwhile to study the continuation of the deformed hyperbolic black holes to Euclidean signature. The rod structure presented in Sec.~\ref{sec5.2} will continue to hold after the analytic continuation $t\rightarrow i\psi$, except that both rods will now be space-like. The two rods are axes of the rotational symmetry generated by $\frac{\partial}{\partial\psi}$ and $\frac{\partial}{\partial\phi}$ respectively. They can be made regular if we impose the following two coordinate identifications: 
\begin{align}
(\psi,\phi)\rightarrow \left(\psi+\frac{2\pi}{\kappa_{\rm 2}},\phi\right)\quad\hbox{and}\quad (\psi,\phi)\rightarrow \left(\psi,\phi+\frac{2\pi}{\kappa_{\rm 1}}\right),
\end{align}
and the metric will be complete in the coordinate range given by the triangular domain. It is then a gravitational instanton, with an underlying manifold $\mathbb{R}^4$. The properties of these gravitational instantons deserve to be studied in more detail.

Ultimately, we hope that the deformed hyperbolic black holes found in this paper, as well as their future generalisations, would find interesting and useful applications via the AdS/CFT correspondence. Since the horizons of these black holes extend to asymptotic infinity $x=y$, the conformal boundary of the space-time {\it will also contain a black hole\/}. It is then possible to use the AdS/CFT correspondence to study the dynamics of a strongly coupled field theory in the background of this boundary black hole. Indeed, this was the programme initiated in \cite{Hubeny:2009ru,Hubeny:2009kz}, and as a first step, one could consider extending this work to the more general class of solutions considered here.

\section*{Acknowledgement}

We would like to thank the referee for enlightening and useful comments. This work was partially supported by the Academic Research Fund (WBS No.: R-144-000-333-112) from the National University of Singapore.

\bigskip\bigskip

{\renewcommand{\Large}{\normalsize}
}

\end{document}